\documentclass[11pt]{article}
\usepackage{graphicx}
\graphicspath{{./figures/}}
\usepackage{appendix}
\usepackage{soul}
\usepackage{verbatim}
\usepackage{comment} 
\usepackage[usenames,dvipsnames]{color}
\usepackage{latexsym,amsmath,amsfonts,amssymb,booktabs}
\usepackage[font=small]{caption}
\usepackage{slashed,upgreek,amscd,cancel,tensor,color}
\usepackage{adjustbox}
\usepackage[numbers,compress,square]{natbib}
\usepackage{epsfig,latexsym}
\usepackage[pdfencoding=auto]{hyperref}
\usepackage{url}
\numberwithin{equation}{section}
\usepackage{doi}
\usepackage{subcaption}
\usepackage[normalem]{ulem}
\definecolor{MyBlue}{rgb}{0.15,0.15,0.70}

\hypersetup{
colorlinks=true,
citecolor=MyBlue,
linkcolor=MyBlue,
urlcolor=MyBlue
}

\input epsf
\usepackage{amssymb}
\usepackage{amsmath}
\usepackage{amsfonts}
\usepackage{upgreek}
\usepackage{latexsym}
\usepackage{stfloats}
\usepackage{afterpage}

\def\re#1{(\ref{#1})}

\newcommand{\fr}{\frac}

\newcommand{\be}{\begin{equation}}
\newcommand{\ee}{\end{equation}}
\newcommand{\beq}{\begin{equation}}
\newcommand{\eeq}{\end{equation}}
\newcommand{\bea}{\begin{eqnarray}}
\newcommand{\eea}{\end{eqnarray}}

\newcommand{\code}[1]{\texttt{#1}}

\newcommand{\bone}{ b_1}
\newcommand{\btwo}{ b_2}
\newcommand{\bthree}{ b_3}
\newcommand{\bfour}{ b_4}

\newcommand{\ctwo}{ c_2}

\newcommand{\cfour}{ c_4}

\definecolor{antiquefuchsia}{rgb}{0.57, 0.36, 0.51}

\usepackage[stable]{footmisc}

\def\dkmu2{\delta K_{\mu \nu}\delta K^{\mu \nu}}
\def\pmu2{  \phi_{\mu \nu}\phi^{\mu \nu}}

\newcommand\bk{{\mathbf{k}}}

\newcommand{\eps}{\epsilon}

\newcommand{\bq}{\mathbf q}

\newcommand\ees{\end{eqnarray}}
\newcommand\bees{\begin{eqnarray}}

\newcommand{\kvec}{{\bf k}}
\newcommand{\qvec}{{\bf q}}
\newcommand{\khat}{\hat{\bf k}}

\newcommand{\cH}{\mathcal{H}}

 \newcommand{\vk}{{\bf k}}
 \newcommand{\vq}{{\bf q}}
 \newcommand{\ta}{\tilde{a}}

\newcommand{\deltam}{\delta}

\newcommand{\Kone}{K_{\delta_g}^{(1)}}
\newcommand{\Ktwo}{K_{\delta_g}^{(2)}}
\newcommand{\Kthree}{K_{\delta_g}^{(3)}}

\newcommand{\CC}{a^{(g)}}
\newcommand{\hh}{g}

\begin{document}
\vspace{0.5cm}

\begin{center}
\Large{\textbf{Constraints on modified gravity from the BOSS galaxy survey}} 

\vspace{.5cm}
L.~Piga$^{1}$, M.~Marinucci$^{2}$, G.~D'Amico$^{1}$, \\ M.~Pietroni$^{1}$,  F.~Vernizzi$^{3}$ and B.~S.~Wright$^{4}$
\vspace{.5cm}
\\ \small{
\textit{$^1$ Department of Mathematical, Physical and Computer Sciences, University of Parma, and INFN Gruppo Collegato di Parma, Parco Area delle Scienze 7/A, 43124, Parma, Italy\\
$^2$ Physics department, Technion, 3200003 Haifa, Israel\\
$^3$ Institut de physique th\' eorique, Universit\'e  Paris Saclay \\ [0.05cm] CEA, CNRS, 91191 Gif-sur-Yvette, France  \\
$^4$ Queen Mary University of London, Mile End Road, London, E1 4NS, United Kingdom }}

\vspace{.2cm}

\end{center}

E-mail: lorenzo.piga@unipr.it, marinucci@campus.technion.ac.il

\vspace{0.5cm}

\begin{abstract}

We develop a pipeline to set new constraints on scale-independent modified gravity, from the galaxy power spectrum in redshift space of BOSS DR12. The latter is modelled using the effective field theory of large-scale structure up to 1-loop order in perturbation theory.
We test our pipeline on synthetic and simulated data, to assess systematic biases on the inferred cosmological parameters due to marginalization and theoretical errors, and we apply it to the normal branch of the DGP model with a $\Lambda$CDM background. When applied to synthetic data and cosmological simulations, we observe biased posteriors due to the strong degeneracy between the nDGP parameter $\Omega_{\rm rc}$  and the primordial amplitude of fluctuations $A_s$. Fixing the latter to the Planck central value, we obtain a posterior distribution with $\Omega_{\rm rc}\lesssim 0.65$ at 95$\%$ C.L., under the assumption of a flat prior on $\log_{10} \Omega_{\rm rc}$. This upper bound, however, depends strongly on the prior on $\Omega_{\rm rc}$. To alleviate this effect, we provide an upper bound based on  the Bayes factor between the nDGP model  and $\Lambda$CDM model, which gives $\Omega_{\rm rc}\lesssim 0.2$ at 95$\%$ C.L..

\end{abstract}

\vspace{1cm}

\tableofcontents

\vspace{.5cm}

\section{Introduction}

The strive to understand the universe has led  to the formulation of the current standard cosmological model: the $\Lambda$CDM model. Besides several currently discussed tensions in the data (see e.g.~\cite{Perivolaropoulos:2021jda} for a recent review), this model  successfully agrees with a wide range of  observations, from the measurements of the late-time accelerated expansion with type Ia supernovae~\cite{SupernovaCosmologyProject:1998vns,SupernovaSearchTeam:1998fmf}, to the cosmic microwave background (CMB) anisotropies~\cite{Planck:2018nkj, Planck:2018vyg} and the baryon acoustic oscillations (BAO) in the galaxy distribution~\cite{SDSS:2005xqv, BOSS:2016wmc}. 
Despite this success, we still lack a deep understanding of the fundamental nature of dark matter and dark energy, leaving open the possibility of alternative scenarios, such as for instance modified gravity models; see e.g.~\cite{Amendola:2016saw}.

An important source of information about the universe comes from large-scale structure (LSS). Future surveys, such as ~DESI~\cite{DESI:2016fyo}, Euclid~\cite{Amendola:2016saw}, and LSST~\cite{LSST:2008ijt}, will map the distribution of galaxies over huge volumes  with unprecedented precision: they will likely become the major source of cosmological information in the coming decades. A major goal of these surveys is the study of the initial conditions and a measurement of neutrino masses, see e.g.~\cite{Ferraro:2022cmj}. But these data will also allow us to perform precision tests of the standard cosmological model and of the theory of general relativity (GR) on large scales, and to exclude many of their extensions.

For this task, it is important to have a  robust understanding of the cosmic web at the scales where the data will be precise enough. Recently, progress has been made in modelling the clustering of galaxies, notably the power spectrum, i.e.~the Fourier transform of the 2-point function, of the galaxy distribution in redshift space. 
In particular, a fairly accurate modelling of the power spectrum can be obtained by using perturbation theory  \cite{Bernardeau:2001qr} up to 1-loop order, supplemented by the renormalization of large-scale fields due to unknown small-scale physics.
In fact, the effect of small-scale physics on correlators can be modeled by the introduction of a set of counterterms \cite{Baumann:2010tm,Pietroni:2011iz,Carrasco:2012cv, Manzotti:2014loa} and bias parameters \cite{McDonald:2009dh,Chan:2012jj,Saito:2014qha,Assassi:2014fva,Senatore:2014eva,Mirbabayi:2014zca,Angulo:2015eqa,Fujita:2016dne} (see \cite{Desjacques:2016bnm} for a review), whose scale dependence can be predicted by symmetries.

This treatment goes under the name of effective field theory of large-scale structure (EFTofLSS) \cite{Baumann:2010tm,Carrasco:2012cv}, which
has been recently used to model the power spectrum at 1-loop for the analysis of the Baryon Oscillation Spectroscopic Survey (BOSS) dataset~\cite{BOSS:2016wmc}, giving constraints on cosmological parameters with a good level of accuracy, see~\cite{DAmico:2019fhj,Ivanov:2019pdj,Chen:2021wdi}.
The results of these and subsequent analyses---e.g.~regarding constraints on neutrino masses~\cite{Colas:2019ret,Ivanov:2019hqk}, the $H_0$ tension~\cite{Philcox:2020vvt,DAmico:2020ods,Ivanov:2020ril}, beyond-$\Lambda$CDM models~\cite{DAmico:2020kxu,DAmico:2020tty}, redshift space distortions (RSD)~\cite{Ivanov:2021fbu, DAmico:2021ymi}---represent an important  step forward in the study of LSS within the EFTofLSS. More recently, the bispectrum---i.e., the Fourier transform of the 3-point function---from BOSS observation has been analyzed using the tree-level~\cite{DAmico:2019fhj, Philcox:2021kcw} and 1-loop EFTofLSS modelling~\cite{DAmico:2022osl}, providing alternative constraints on primordial non-gaussianities~\cite{DAmico:2022gki,Cabass:2022wjy,Cabass:2022ymb}.

In this paper we show that current and future LSS data can be used to put reliable constraints on modified gravity as well. 
Specifically, we develop a pipeline to analyze the galaxy power spectrum  in redshift space, that can be used to test general scale-independent extensions of the $\Lambda$CDM model. Then we apply it to BOSS data and constrain specifically the so-called normal branch\footnote{The self-accelerating branch~\cite{Deffayet:2000uy} of the DGP model is unstable~\cite{Luty:2003vm,Charmousis:2006pn,Koyama:2007za}.} of the Dvali-Gabadadze-Porrati (DGP) model~\cite{Dvali:2000hr}, or nDGP for short. 
Despite introducing a single additional parameter with respect to $\Lambda$CDM, it incorporates  many interesting cosmological features: modifications in the background evolution, in the strength of the gravitational clustering and in the dynamics of the fluctuations, at linear order and beyond. For these reasons, this model is one of the most studied modification of GR and has been implemented in several $N$-body simulations, see e.g.~\cite{Hernandez-Aguayo:2020kgq} and references therein.

In the original DGP model, the universe is described by a 4d brane embedded in a 5d Minkowski spacetime. 
The cross-over scale between the 5d and 4d behaviour is given by the length scale $r_c$, which represents the fundamental extra parameter of this model, with GR being smoothly recovered by taking the cross-over scale much larger than the current Hubble scale $H_0^{-1}$, i.e., for $H_0 r_c \gg 1$. Constraints on this parameter are traditionally expressed in terms of the dimensionless quantity \cite{Lombriser:2009xg}
\be
\label{Orcdef}
\Omega_{\rm rc} \equiv \frac{1}{4 r_c^2 H_0^2}.
\ee
For instance, ref.~\cite{Lombriser:2009xg} 
puts an upper bound to the nDGP model of $\Omega_{\rm rc} < 0.020$ (95\%~C.L.) from WMAP and other available CMB data (including  ISW-galaxy correlation), supernovae and measurements of $H_0$, assuming a cosmological constant in the background and a modified expansion history without spatial curvature. A more recent analysis with the same hypothesis was performed in \cite{Xu:2013ega} including CMB data from Planck, supernovae and BAO. 

In contrast with these analyses, in the following we consider a version of the nDGP model where the background expansion history exactly reproduces the $\Lambda$CDM one in a spatially-flat universe. This is obtained by considering, instead of a cosmological constant,  a dark energy component whose background dynamics exactly compensates the modified background evolution of the nDGP model \cite{Schmidt:2009sv}.  
Deviations from $\Lambda$CDM take place only in the late-time evolution of perturbations, so that the model is weakly constrained by supernovae, sensitive only to the recent background expansion history. Similarly, it is weakly constrained by CMB data, whose dependence on late-time perturbations is only through the small integrated Sachs-Wolfe effect and lensing. Focusing on a nDGP model with $\Lambda$CDM background expansion allows us to probe the constraining power of LSS data, which are the most sensitive to the late-time evolution of perturbations.

Previous constraints on the nDGP model obtained in this setting are much weaker: for instance in~\cite{2013MNRAS.436...89R} it was found $\Omega_{\rm rc}\lesssim 40$, using measurements of the monopole and the quadrupole of the correlation function from SDSS DR7 data, with fixed $H_0$. 
More recently, Ref.~\cite{Barreira:2016ovx} used the values of the parameter combination $f\sigma_8$ estimated from the BOSS DR12 data \cite{BOSS:2016off}  to set an upper bound of $\Omega_{\rm rc}\lesssim 0.25$ at 95\% C.L. This constraint was obtained by setting tight gaussian priors on both  $\Omega_{\mathrm m0}$ and $\sigma_{8}^{\Lambda CDM}\mathrm{(z= 0)}$, corresponding to $1\sigma$ Planck constraints assuming flat $\Lambda$CDM.

As we will see below, in this work we obtain $\Omega_{\rm rc}\lesssim 0.65$ at 95$\%$ C.L. from the BOSS dataset. This constraint can be obtained only after fixing the primordial amplitude of fluctuations $A_s$ derived from Planck's measurements. Relaxing $A_s$, the bound obtained is much looser and biased because the effect of modified gravity on the growth of structures is very degenerate with  $A_s$. In principle, this degeneracy could be broken by combining data from different redshift bins, as the effect of $\Omega_{\rm rc}$ is time dependent. While this could be an effective strategy for future surveys such as DESI and Euclid, this is not the case for the BOSS data, as the two redshift bins analyzed in this paper are very close.  

The constraint presented above depends on the choice of prior for the $\Omega_{\rm rc}$ parameter. For this reason we also provide a prior-independent constraint on the nDGP parameter based on the so-called Bayes factor~\cite{Gordon:2007xm}. This procedure is particularly useful for constraining parameters with a non-gaussian posterior. With this approach we obtain the upper value $\Omega_{\rm rc}\lesssim 0.2$ at 95$\%$ C.L..

We model the galaxy redshift-space power spectrum using the EFTofLSS implemented in \code{PyBird} ~\cite{DAmico:2020kxu}, a fast Python code for the evaluation of the 1-loop power spectrum, developed for $\Lambda$CDM and modified here to include scale-independent modifications of gravity. We stress that, although our analysis is restricted to the nDGP model alone, it can be straightforwardly extended to  other scale-independent models, such as those described by the general single-field framework of the EFT of dark energy \cite{Creminelli:2008wc,Gubitosi:2012hu,Bloomfield:2012ff,Gleyzes:2013ooa,Bloomfield:2013efa} extended beyond linear order \cite{Cusin:2017mzw,Cusin:2017wjg}. This opens the possibility of putting tighter constraints on a variety of dark energy and modified gravity scenarios by combining Planck data with future surveys that will scan larger volumes and will contain higher number of galaxies than SDSS-III.

The paper is structured as follows. In sec.~\ref{sec:PTMG} we calculate the 1-loop power spectrum for galaxies in redshift space for a generic scale-independent modified gravity model. Section~\ref{sec:PB_nDGP} describes the implementation and the validation of the nDGP model  in \code{PyBird}. 
Before proceeding  with the BOSS analysis, we devote sec.~\ref{sec:pipeline} to reveal the presence of projections effects, i.e.~systematic biases on the 1-d  posteriors due to  marginalization over the other parameters. We do that by testing our pipeline on synthetic data
and on simulations with known cosmological parameters and error specifications adapted to the two BOSS redshift samples. In particular, we use the so-called ``PT Challenge'' simulations, a set of high-resolution mocks mimicking the SDSS-III BOSS galaxy samples but covering a hundred times larger cumulative volume, which were employed in a blinded challenge consisting in inferring cosmological parameters from the power spectrum multipoles~\cite{Nishimichi:2020tvu} (see sec.~\ref{sec: sims} for more details). Testing on simulations allows us also to study  theory errors and the largest wavenumber $k_{\rm max}$ that we can reliably use within the effective theory.
In sec.~\ref{sec:BOSS} we present the result of our analysis on the BOSS data and discuss the bounds on $\Omega_{\rm rc}$. We conclude in sec.~\ref{sec:concl}.
We relegate several discussions to the appendix. In particular, we display the full time-dependent functions of the nDGP scenario in app.~\ref{appendixTD}.
Appendix~\ref{app:map} presents a comparison of the bias expansion used in this work with the one of the LSS bootstrap introduced in~\cite{DAmico:2021rdb} that extends to modified gravity models. Moreover,  we discuss how to choose the initial conditions for the linear growth factor in app.~\ref{app:IC_nDGP}. 

We note that a  pipeline analogous to the one presented here, based on the EFTofLSS, 
has been recently developed and used \cite{Carrilho:2022mon} to constrain the dark scattering model of interacting dark
energy \cite{Simpson:2010vh} with BOSS data.

\section{Galaxy clustering in modified gravity}
\label{sec:PTMG}

In this section we work out the expression of the galaxy power spectrum in redshift space for generic {\em scale-independent} modified gravity models. 
In sec.~\ref{sub2.1}, following earlier works, see e.g.~\cite{Taruya:2014faa,Cusin:2017wjg}, we first introduce the evolution equations of dark matter in modified gravity, focusing on standard perturbation theory without counterterms. The exact solutions to these equations in perturbation theory are derived in sec.~\ref{sub2.2}. We then discuss the bias expansion in redshift space in sec.~\ref{sec:BT_MG}, following \cite{Donath:2020abv}. In this reference, the bias expansion was derived in the  exact time dependent case (i.e.~without assuming the commonly used Einstein-de Sitter approximation) for the $\Lambda$CDM model. In \cite{DAmico:2021rdb}, some of us showed that this expansion also holds for scale-independent modified gravity models sharing the same symmetries as $\Lambda$CDM. Finally, in the last subsection, \ref{bootPS}, we write down the 1-loop power spectrum, including all counterterms from the EFTofLSS used in \code{PyBird}.

\subsection{Dark matter dynamics}
\label{sub2.1}

We consider a perturbed spatially-flat Friedmann-Lema\^itre-Robertson-Walker metric in the Newtonian gauge, focusing on scalar perturbations, i.e.
\be
ds^2 = - (1+ 2 \Phi) dt^2 + a^2(t) (1- 2 \Psi) d{\bf x}^2 \;.
\ee
We work in the Jordan frame, where matter is minimally coupled to the gravitational metric. In this case test particles follow geodesics and the dark matter fluid is  described by the standard continuity and Euler equations 
\begin{align}  \label{finalconteq}
& \dot \delta + a^{-1} \partial_i \left( (1+ \delta ) v^i \right) = 0 \ , \\
&  \dot v^i + H v^i + \frac1{a} v^j \partial_j v^i   + \frac{1}{a}  \partial_i \Phi   = - \frac{1}{a \rho_{\rm m}} \partial_j \tau^{ij} \ ,    \label{finaleulereq} \end{align}
where $\rho_{\rm m}$ is the dark matter energy density, with background value $\bar \rho_{\rm m}$, $\delta \equiv \rho_{\rm m}/\bar \rho_{\rm m}-1$ and $v^i$ are respectively the energy density contrast and the velocity of dark matter, a dot denotes the derivative with respect to the cosmic time $t$ and $H \equiv \dot a/a$ is the Hubble rate. Following \cite{Baumann:2010tm,Pietroni:2011iz,Carrasco:2012cv}, we have written down the smoothed continuity and Euler equations: the right-hand side of eq.~\eqref{finaleulereq} is the effective stress-energy tensor describing how the short modes affect the dynamics of the long modes resulting from this smoothing procedure. See \cite{Cusin:2017wjg} for a generalization to modified gravity models of this smoothing procedure and stress-energy tensor.

In GR, one closes these two equations with the Poisson equation. In modified gravity, however, the Poisson equation no longer holds. In order to proceed, we must thus discuss how to relate $\Phi$ to the fluid variables. To do so, in the following we specialize to scale-independent models, i.e.~we assume that the mass of the scalar field responsible for modifying gravity is much smaller than the fundamental frequency of the survey. This is typically the case for Horndeski theories \cite{Horndeski:1974wa,Deffayet:2011gz}, where higher-order derivative terms can lead to self-acceleration \cite{Luty:2003vm,Nicolis:2008in} and display Vainshtein screening \cite{Vainshtein:1972sx} around overdense regions. 
In this case, one can use the full field equations involving both $\Phi$ and $\Psi$(\footnote{In these theories $\Phi$ and $\Psi$ are not necessarily the same.}) to express the Laplacian of $\Phi$ in terms of the density contrast. This expression  contains terms linear in $\delta$, as in the Poisson equation, but in general   there are also higher-order terms. Since we are interested  in computing the 1-loop power spectrum, we will only consider  terms up to third order in $\delta$. 

In summary, in a generic model one obtains (see e.g.~\cite{Cusin:2017mzw,Cusin:2017wjg})
 \begin{align}
\label{sol_NL1}
\frac{\partial^2 \Phi}{H^2 a^2} = & \   \frac{3 \, \Omega_{{\rm m},a} }{2}  \,  \nu_{ } \,  \deltam+  \left( \frac{3\, \Omega_{{\rm m},a}}{2}  \right)^2 \nu_{2}  \left[\deltam^2-\left(\partial^{-2}{\partial_i\partial_j}\deltam\right)^2\right]  \\
&+\left(  \frac{3\, \Omega_{{\rm m},a}}{2} \right)^3 \nu_{22}\left[\deltam-\left(\partial^{-2} {\partial_i\partial_j} \deltam\right) \partial^{-2} {\partial_i\partial_j} \right]\left[\deltam^2-\left(\partial^{-2}{\partial_k\partial_l} \deltam\right)^2\right]   \nonumber  \\
&+\left(  \frac{3\, \Omega_{{\rm m},a}}{2}  \right)^3 \nu_{3}\left[\deltam^3-3\deltam\left(\partial^{-2}{\partial_i\partial_j} \deltam\right)^2+2 (\partial^{-2}{\partial_i\partial_j} \deltam)( \partial^{-2}{\partial_k\partial_j}\deltam )( \partial^{-2}{\partial_i\partial_k}\deltam )\right]    + {\cal O} (\delta^4)\,, \nonumber
\end{align}
where 
\be
\label{Omdef}
\Omega_{{\rm m},a} \equiv \frac{\bar \rho_{\rm m} }{3 M^2 H^2} \; 
\ee
is the {\em time-dependent} dark matter energy density in critical units and  $M $ is the effective Planck mass, which in general can  depend on time.
The functions $\nu ( a )$, $\nu_{2} ( a )$, $\nu_{22} ( a )$, and $\nu_{3 }(a)$ parametrize the time-dependent amplitude of the higher order terms.  They can be related to the original parameters of the modified gravity model at hand\footnote{The derivation of eq.~\eqref{sol_NL1} is explicitly laid out in \cite{Cusin:2017mzw} for Horndeski models described in terms of the EFT of dark energy. There, one also finds the relations between $\nu$, $\nu_{2} $, $\nu_{22} $, and $\nu_{3 }$ and the EFT of dark energy parameters, expressed in terms of the Horndeski Lagrangian. The analogous expression for the nDGP model is derived in \cite{Bose:2018orj}.} but from the viewpoint of the LSS equations, they are simply  functions of time.\footnote{Note that this is only valid on length scales above the nonlinear scale where $\delta \sim 1$ and above the Vainshtein screening \cite{Vainshtein:1972sx} scale where scalar field fluctuations enter the nonlinear regime, see \cite{Cusin:2017mzw}.} In general relativity $M = M_{\rm Pl} = (8 \pi G)^{-1/2}$, $\nu=1$,  $\nu_{2}=\nu_{3}=\nu_{22}=0$ and we recover  the standard Poisson equation.
Note that $\Psi$ does not directly appear in the fluid equations in the non-relativistic limit considered here and we do not need it  for this study.

At linear order one can neglect the higher-order terms in eqs.~\eqref{finalconteq}--\eqref{sol_NL1} and use these three equations to write a single (second-order in time) equation for the linearized density contrast $\delta^{(1)}(a)$. Its time dependence is  captured by the growth factors $D(a)$, defined such that 
\be
\delta^{(1)}(a) = D(a) \delta^{(1)}(a_{\rm in}) \;, 
\ee
satisfying  
			\be\label{growth}
			\frac{d^2 D(a) }{d\ln a^2}+\bigg(2+\frac{d\ln H}{d\ln a}\bigg)\frac{d D(a)}{d\ln a}-\fr{3}{2} \nu  (a) \Omega_{{\rm m},a} (a)D(a)=0 \; ,
			\ee
and normalized to unity  at some initial time deep in matter domination era, $a=a_{\rm in}$.
Assuming a $\Lambda$CDM background expansion rate, the difference with the standard $\Lambda$CDM case is captured by the function $\nu (a)$ in the last term, which  modifies the strength of the gravitational interaction. 
In the following we assume that $\nu  \to 1$ (i.e., GR is recovered) at early time. Since in this limit $\Omega_{{\rm m},a} \to 1$, at early time the solutions of this equation are the usual growing and decaying solutions in matter domination, i.e.~$D_+ \propto a$ and $D_-\propto a^{-3/2}$, respectively.  
We can thus label the  late-time solution as growing (decaying) if at early times it grows as $a$ (decays as $a^{-3/2}$). The Green's functions of this equation are discussed in the next subsection.

To proceed, we also define the linear growth rate,
\be
\label{deff}
f \equiv \frac{d \ln D}{d \ln a} \;.
\ee
It satisfies the equation
\be
\label{ffeq}
\frac{d f}{d \ln a} + f^2 + \left(2 + \frac{d \ln H}{d \ln a}  \right) f - \frac{3}{2} \nu  \Omega_{{\rm m},a} =0 \;,
\ee
that can be straightforwardly derived from eq.~\eqref{growth}. We define $f_+$ and $f_-$ by eq.~\eqref{deff} with $D=D_+$ and $D=D_-$, respectively.

It is now convenient to go to Fourier space and study the dynamics of $\delta$ and $v^i$ as a function of the scale factor. We neglect momentarily the stress-energy tensor on the right-hand side of eq.~\eqref{finaleulereq}. Its effect will be discussed in sec.~\ref{bootPS}.
Thus, we
define the conformal Hubble rate $\cH \equiv a H$ and use a prime to denote the derivative with respect to  $a$. Moreover, we also introduce the rescaled velocity divergence $\theta$, defined as
\be
\label{thetaV}
\theta \equiv - \frac{\partial_i v^i }{ f_+ \cH } \;.  
\ee
Neglecting vorticity modes, the dynamics of  $\delta$ and $\theta$ are now described by
\begin{align} 
a \, \delta_\kvec ' (   a ) - f_+ \theta_\kvec (  a ) & =  \int_{\kvec_1, \kvec_2}  ( 2 \pi )^3 \delta_D ( \kvec - \kvec_{12}  ) \nonumber  \\
& \hspace{0.1in} \times f_+ \alpha ( \kvec_1 , \kvec_2 ) \theta_{\kvec_1 } (  a ) \delta_{\kvec_2} (   a) \label{conteq1} \;, \\
a \, \theta_ \kvec ' (  a ) -f_+ \theta_\kvec (   a ) + \frac32 \frac{\nu   \,\Omega_{{\rm m},a} }{f_+} \theta_\kvec (   a ) + \frac{1}{f_+} \frac{k^2}{\cH^2}  \Phi_\kvec ( a)  & =  \int_{\kvec_1,\kvec_2}  ( 2 \pi )^3 \delta_D ( \kvec - \kvec_{12}  ) \nonumber   \\
& \hspace{0.1in} \times f_+ \beta ( \kvec_1 , \kvec_2 ) \theta_{ \kvec_1} (  a ) \theta_{\kvec_2} (  a) \;, \label{euleq1}
\end{align} 
where  to derive the second equation we have used eq.~\eqref{ffeq}. Moreover,
$\alpha ( \kvec_1 , \kvec_2 )$ and $\beta ( \kvec_1 , \kvec_2 )$ are the standard dark matter interaction vertices,
\begin{align}  
\alpha ( \kvec_1 , \kvec_2 ) = 1 + \frac{\kvec_1 \cdot \kvec_2}{k_1^2}  \hspace{.3in} \text{and} \hspace{.3in}
\beta( \kvec_1 , \kvec_2 )  = \frac{ | \kvec_1 + \kvec_2 |^2 \kvec_1 \cdot \kvec_2}{2 k_1^2 k_2^2}    \label{betadef2new} \ ,
 \end{align}
 and we have used the notation $\int_{\kvec_1, \ldots, \kvec_n } \equiv \int \frac{ d^3 k_1}{( 2 \pi )^3} \cdots \int \frac{ d^3 k_n}{( 2 \pi )^3}$ and $\kvec_{1\ldots n} \equiv \kvec_1  + \ldots + \kvec_n$.

In Fourier space, the generalized Poisson equation \eqref{sol_NL1} reads
\begin{align} \label{newd2phi}
- \frac{ k^2}{\cH^2 } \Phi_\kvec ( a ) & =  \nu   \frac{3 \, \Omega_{{\rm m},a}   }{2 } \, \delta_ \kvec (a  ) \\
& \hspace{-.4in} + \nu_{2}     \left( \frac{ 3 \, \Omega_{{\rm m},a} }{2  } \right)^2     \int_{\kvec_1, \kvec_2} ( 2 \pi )^3 \delta_D ( \kvec - \kvec_{12} ) \,  \gamma ( \kvec_1 , \kvec_2  ) \delta_{ \kvec_1} (a ) \delta_{\kvec_2} ( a) \nonumber \\
& \hspace{-.4in} +  \nu_{3}      \left( \frac{ 3 \, \Omega_{{\rm m},a} }{2  } \right)^3   \int_{\kvec_1, \kvec_2, \kvec_3}  ( 2 \pi)^3 \delta_D ( \kvec - \kvec_{123} ) \gamma_3 ( \kvec_1 , \kvec_2 , \kvec_3 ) \delta_{\kvec_1}  ( a) \delta_{\kvec_2}( a ) \delta_{\kvec_3 } (  a) \nonumber \\
& \hspace{-.4in} +   \nu_{22}      \left( \frac{ 3 \, \Omega_{{\rm m},a} }{2  } \right)^3    \int_{\kvec_1, \kvec_2,\qvec_1, \qvec_2}  ( 2 \pi)^3 \delta_D ( \kvec - \kvec_{12}) ( 2 \pi)^3 \delta_D ( \kvec_2 - \qvec_{12}) \nonumber \\
& \hspace{1.4in} \times \gamma ( \kvec_1 , \kvec_2 ) \gamma ( \qvec_1 , \qvec_2 ) \delta_{\kvec_1} ( a ) \delta_{\qvec_1}( a) \delta_{\qvec_2} (  a) \nonumber \ ,
\end{align}
where the new kernels inside the integrals are given by
\be
\label{gammas}
\begin{split}
\gamma ( \kvec_1 , \kvec_2  ) & =   1 -  \big( \khat_1 \cdot \khat_2 \big)^2 \;,  \\
\gamma_3 ( \kvec_1 , \kvec_2 , \kvec_3 ) & =    1  + 2 \big( \khat_1 \cdot \khat_2\big)  \, \big( \khat_1 \cdot \khat_3 \big) \, \big( \khat_2 \cdot \khat_3 \big)    - \big( \khat_1 \cdot \khat_3 \big)^2  -  \big( \khat_2 \cdot \khat_3 \big)^2   - \big( \khat_1 \cdot \khat_2 \big)^2  \ . 
\end{split}
\ee

For completeness, we have included above  the cubic vertex proportional to $\nu_{3}$. However, since in the PS at 1-loop it enters as $\gamma_3 (\kvec, \qvec, -\qvec) =0$ \cite{Cusin:2017wjg} it does not contribute to the power spectrum at one loop and we will discard it. Therefore, removing this term the perturbation equations above  become
		\begin{align}
		\label{eq:master1}
		a\delta_{\vk}' (a)-f_{+}(a)\theta_{\vk}(a)&= S^\delta_{\vk}(a)\; ,\\
		a\theta_{\vk}'(a)-f_{+}(a)\theta_{\vk}(a)+\fr{3}{2}  \frac{\nu (a) \Omega_{{\rm m},a}(a) }{f_{+}(a) } \left(\theta_{\vk}(a)-\delta_{\vk}(a) \right)&= S^\theta_{\vk}(a) \;,
		\label{eq:master2}
		\end{align}
		with
\begin{align}
S_{\vk}^\delta &= (2\pi)^{3}\int_{ \kvec_1 \kvec_2}   \delta_{D}(\vk-\vk_{12})f_{+}  \alpha(\vk_1,\vk_2)\theta_{\vk_1}\delta_{\vk_2} \;, \\
S_{\vk}^\theta &= (2\pi)^{3} \int_{ \kvec_1 \kvec_2}  \delta_{D}(\vk-\vk_{12})  \left[ f_{+} \beta(\vk_1,\vk_2)\theta_{\vk_1}\theta_{\vk_2} + \frac{\nu_{2}}{f_+} \left( \frac{3 \Omega_{{\rm m},a}}{2}  \right)^2  \gamma (\vk_1,\vk_2)\delta_{\vk_1}\delta_{\vk_2} \right] \nonumber \\
		& + (2\pi)^{3} \int_{\kvec_1 \kvec_2 \qvec_1 \qvec_2}  \delta_{D}(\vk_2-\vq_{12} ) \delta_{D}(\vk-\vk_{12})  \frac{\nu_{22}}{f_+} \left( \frac{3 \Omega_{{\rm m},a}}{2}  \right)^3 \gamma(\vq_1,\vq_2) \gamma (\vk_1,\vk_2) \delta_{\vk_1} \delta_{\vq_1}\delta_{\vq_2} \; .
\end{align}
Moreover, for later purposes it is convenient to define the  symmetrized $\alpha$ as $\alpha_s(\vk_1,\vk_2) = \fr{1}{2}(\alpha(\vk_1,\vk_2)+\alpha(\vk_2,\vk_1))$ and notice that
\be
\gamma ( \kvec_1 , \kvec_2  )  = \alpha_s ( \kvec_1 , \kvec_2  )  - \beta ( \kvec_1 , \kvec_2  ) \;.
\ee

\subsection{Perturbative solutions for dark matter}
\label{sub2.2}

Following \cite{Lewandowski:2016yce}, to construct the higher-order solutions to  eqs.~(\ref{eq:master1}) and (\ref{eq:master2}), we need the Green's functions. For general scale-independent modified gravity models, where the deviations from GR are captured at linear order by the time-dependent function $\nu $, these are defined as  
			\begin{align}
			&a \frac{d G^{\delta}_{\sigma}(a,\ta)}{da}-f_{+}G^{\theta}_{\sigma}(a,\ta)=\lambda_{\sigma}\delta_D(a-\ta), \label{Green} \\
			&a \frac{d G^{\theta}_{\sigma}(a,\ta)}{da}-f_{+} G^{\theta}_{\sigma}(a,\ta)+\fr{3}{2} \frac{\nu  \Omega_{{\rm m},a}}{f_{+}}\bigg(G^{\theta}_{\sigma}(a,\ta)-G^{\delta}_{\sigma}(a,\ta)\bigg)=(1-\lambda_{\sigma})\delta_D(a-\ta),
			\label{Greenf}
			\end{align}
			where $\lambda_1 = 1$ and $\lambda_2 = 0$. 

			Explicitly, they are given by
			\begin{align}
			&G^{\delta}_1(a,\ta)=\frac{1}{\ta W(\ta)}\bigg(\frac{d D_{-}(\ta)}{d\ta}D_{+}(a)-\frac{d D_{+}(\ta)}{d\ta}D_{-}(a)\bigg) {\Theta}(a-\ta) \label{gdelta} \ ,\\
			&G^{\delta}_2(a,\ta)=\frac{f_{+}(\ta)/\ta^2}{W(\ta)}\bigg(D_{+}(\ta)D_{-}(a)-D_{-}(\ta)D_{+}(a)\bigg){\Theta}(a-\ta) \ , \\
			&G^{\theta}_1(a,\ta)=\frac{a/\ta}{f_{+}(a)W(\ta)}\bigg(\frac{d D_{-}(\ta)}{d\ta}\frac{d D_{+}(a)}{d a}-\frac{d D_{+}(\ta)}{d\ta}\frac{d D_{-}(a)}{d a}\bigg) {\Theta}(a-\ta) \ ,\\
			&G^{\theta}_2(a,\ta)=\frac{f_{+}(\ta)a/\ta^2}{f_{+}(a)W(\ta)}\bigg(D_{+}(\ta)\frac{d D_{-}(a)}{d a}-D_{-}(\ta)\frac{d D_{+}(a)}{d a}\bigg) {\Theta}(a-\ta) \ , \label{gtheta}
			\end{align}
			where $W(\ta)$ is the Wronskian of $D_+$ and $D_-$, 
			\be
			\label{wrk}
			W(\ta)=\frac{dD_{-}(\ta)}{d\ta}D_{+}(\ta)-\frac{d D_{+}(\ta)}{d\ta}D_{-}(\ta) \ ,
			\ee
			$\Theta (a-\tilde a)$ is the Heaviside step function and we impose the boundary conditions \begin{align} \label{bound}
			& G^{\delta}_\sigma(a,\tilde a) = 0 \quad \quad \text{and} \quad\quad G^{\theta}_\sigma(a, \tilde a)=0 \quad \quad \text{for} \quad \quad \tilde a > a \ , \\
			&G^\delta_\sigma ( \tilde a , \tilde a ) = \frac{\lambda_\sigma}{\tilde a} \quad \hspace{.06in} \text{and} \hspace{.2in} \quad G^{\theta}_{\sigma} ( \tilde a , \tilde a ) = \frac{(1 - \lambda_\sigma)}{\tilde a} \; . \label{bound2}
			\end{align}

We write the dark matter density contrast and velocity divergence in a perturbative expansion of the form
		\be \label{deltavexp}
		\delta_{\vk}(a)=\sum^{\infty}_{n=1}\delta_{\vk}^{(n)}(a) \quad \textmd{and} \quad \theta_{\vk}(a) =\sum^{\infty}_{n=1} \theta_{\vk}^{(n)}(a),
		\ee
		which allows us to solve equations \eqref{eq:master1} and \eqref{eq:master2} order by order.  
		The perturbative solution of these equations can then be written as an integral over time-dependent momentum kernels,
		\be
		\begin{split}
		\label{eq:kernelform}
		\delta^{(n)}_{\vk}(a) &=\int\frac{d^3q_1}{(2\pi)^{3}} \ldots \frac{d^3q_n}{(2\pi)^{3}}(2\pi)^{3}\delta_{D}(\vk-\vq_{1n})K_\delta^{(n)}(\vq_1,\ldots ,\vq_n,a)\delta^{(1)}_{\vq_1}(a)\ldots \delta^{(1)}_{\vq_n}(a) \;, \\ 
		\theta^{(n)}_{\vk}(a)&=\int\frac{d^3q_1}{(2\pi)^{3}}\ldots \frac{d^3q_n}{(2\pi)^{3}}(2\pi)^{3}\delta_{D}(\vk-\vq_{1n})K_\theta^{(n)}(\vq_1,\ldots ,\vq_n,a)\delta^{(1)}_{\vq_1}(a)\ldots \delta^{(1)}_{\vq_n}(a) \; .
		\end{split}
		\ee
	 Up to third order, the kernels are given by \cite{Donath:2020abv},\footnote{In \cite{Donath:2020abv}, eqs.~\re{kernels1}--\re{kernels3} are derived  assuming $\Lambda$CDM but their validity extends also to other models respecting the same symmetries, see~\cite{DAmico:2021rdb}.}
		\bea\label{kernels1}
		K_\lambda^{(1)}(\vq_1,a)&=& 1 \;, \\
		\label{kernels2}
		K_\lambda^{(2)}(\vq_1,\vq_2,a)&=&\alpha_s(\vq_1,\vq_2){\cal G}^{\lambda}_{1}+\beta(\vq_1,\vq_2){\cal G}^{\lambda}_{2} \;, \\
		\label{kernels3}
		K_\lambda^{(3)}(\vq_1,\vq_2,\vq_3,a)&=&\alpha^{\sigma}(\vq_1,\vq_2,\vq_3) {\cal U}^{\lambda}_{\sigma} +\beta^{\sigma}(\vq_1,\vq_2,\vq_3) {\cal V} ^{\lambda}_{\sigma2} + \gamma^{\sigma}(\vq_1,\vq_2,\vq_3){\cal V}^{\lambda}_{\sigma1} \;,
		\eea
	where repeated $\sigma \in \{1,2\}$ are summed over and $\lambda \in \{\delta,\theta\}$. The six momentum kernels at third order $\{\alpha_\sigma,\beta_\sigma,\gamma_\sigma\}$ are products of $\alpha_s$ and $\beta$, while  $\{{\cal G}^{\lambda}_1,{\cal G}^{\lambda}_2, {\cal U}^{\lambda}_{\sigma}, {\cal V}^{\lambda}_{\sigma\tilde{\sigma}}\}$, where $\tilde{\sigma} \in \{1,2\}$, are time-dependent functions resulting from equations (\ref{eq:master1}) and (\ref{eq:master2}). All these functions are explicitly given in app.~\ref{appendixTD}. Moreover, in app.~\ref{subapp1} we discuss the relation between this expansion and the (equivalent) one derived using the bootstrap approach \cite{DAmico:2021rdb}.

\subsection{Biased tracers in redshift space}	
\label{sec:BT_MG}

Galaxies are biased tracers of the long wavelength dark matter field distribution (see~\cite{Desjacques:2016bnm} for a review). As such, their density distribution $\delta_g$ can be related to the linear dark matter density distribution $\delta^{(1)}$ and its derivatives (see e.g.~\cite{Chan:2012jj,Saito:2014qha,Assassi:2014fva,Mirbabayi:2014zca,Senatore:2014eva,Angulo:2015eqa,Fujita:2016dne}). This relation is encoded by a bias expansion given in terms of the kernels $K_{\delta_g}^{(n)}$,
defined by
\be
\delta_{{ g},\bk}^{(n)}(\eta) = \int\frac{d^3\bq_1}{(2\pi)^3}\dots\int\frac{d^3\bq_n}{(2\pi)^3}(2\pi)^3 \delta_D(\bk - \bq_{1\dots n})K_{\delta_g}^{(n)} (\bq_1,\dots,\bq_n,\eta)\delta_{\bq_1}^{(1)}(\eta)\dots\delta_{\bq_n}^{(1)}(\eta)\,.
\label{PT_tr}
\ee

In what follows we use the same bias expansion and kernels introduced in \cite{DAmico:2019fhj} for a $\Lambda$CDM cosmology using the Einstein-de Sitter (EdS) approximation  and extended to an exact time-dependent evolution in \cite{Donath:2020abv}.  It was shown in  \cite{DAmico:2021rdb}, using the so-called bootstrap approach, that this bias expansion does not restrict to $\Lambda$CDM but  can be straightforwardly applied to a large class of modified gravity models, i.e.~all those models that share the same symmetries with $\Lambda$CDM. It can be thus  applied to the nDGP model and to all single-field models within the Horndeski class.

Using this approach, one can show that the kernels up to third order can be expressed in terms of seven perturbative bias coefficients. 
However, for the calculation of the 1-loop power spectrum the third-order kernel appears with a particular combination of momenta, $K_{\delta_g}^{(3)}(\bq,-\bq,\bk;a)$, from which we subtract its $|\bq|/|\bk|\to \infty$  limit. 
This reduces the effective number of bias parameters that enter the calculation to four, that we  denote by $\bone$, $\btwo$, $\bthree$, and $\bfour$.\footnote{In the EdS-approximation limit,   this bias expansion is equivalent to the one introduced in \cite{DAmico:2019fhj}. } See app.~\ref{subapp2} for details.

The relevant terms then are  given by 
\begin{align}
  &\label{Kdg1} K_{\delta_g}^{(1)}(\bk;a) = \bone\,,\\
  &\label{Kdg2} K_{\delta_g}^{(2)}(\bq_1,\bq_2;a) = (- \bone + \btwo + \bfour) + \bone \beta(\bq_1,\bq_2)+ \left(\bone - \frac27 \btwo \right) \gamma(\bq_1,\bq_2)   \,,\\
  &\!\! \left. K_{\delta_g}^{(3)}(\bq_1,\bq_2,\bq_3;a)\right|_{\rm sub} =
\frac{\bone}{3} O_{\beta\beta}(\bq_1,\bq_2,\bq_3)+\frac{1}{3}\left(\frac{g(a) \bone}{2}+\frac{\bthree}{21} \right)O_{\gamma\beta}(\bq_1,\bq_2,\bq_3)\nonumber\\
 &\qquad\qquad \qquad\qquad\quad\; +\frac{1}{3}\left( \frac{g(a) \bone}{2}-\frac{\bthree}{21} \right)\left(O_{\gamma\gamma}(\bq_1,\bq_2,\bq_3) +\frac{1}{2}O_{\gamma\alpha_a}(\bq_1,\bq_2,\bq_3)\right)+ {\rm cyclic},\label{Kdg3}
\end{align} 
where we have defined,
\be
\label{OO}
O_{X Y}(\bq_1,\bq_2,\bq_3)\equiv \,X(\bq_1,\bq_2)\, Y(\bq_{12},\bq_3)\,,
\ee
where $X,Y$ are the kernels $\beta$, $\gamma$ and $\alpha_a$, with
\be 
\alpha_a(\bq_{1},\bq_2)\equiv \alpha(\bq_{1},\bq_2)-\alpha(\bq_{2},\bq_1)\,,
\ee
and $g$ is the {\em tracer-independent} time-dependent function, 
\beq
g(a) = 2\, {\cal G}_1^\delta(a)-1\,,
\eeq
as shown in app.~\ref{subapp1}.

To derive the  density contrast of the galaxies in redshift space, we can follow the usual procedure 
described in \cite{Scoccimarro:1999ed,Senatore:2014vja}, which gives 
\be
\label{expansion_z2}
\delta_{g,s}^{(n)}(\bk;\eta) \equiv \int\frac{d^3\bq_1}{(2\pi)^3} \dots\int\frac{d^3\bq_n}{(2\pi)^3} (2\pi)^3\delta_D\left(\bk - \bq_{1\dots n}\right)Z^{(n)}(\bq_1,\dots,\bq_n;\eta)\delta^{(1)}_{\bq_1}(\eta)\dots\delta^{(1)}_{\bq_n}(\eta)\,,
\ee
where
\begin{align}
Z^{(1)}(\bq_1) = \ & b_1 + f\mu_1^2 \,, \label{Zone}\\ 
Z^{(2)}(\bq_1,\bq_2) = \ & K^{(2)}_{\delta_g}(\bq_1,\bq_2) + f\mu_{k}^2 K^{(2)}_{\theta}(\bq_1,\bq_2)\nonumber\\
& + f\mu_{k}k\left[\frac{\mu_1}{q_1} K_\theta^{(1)} (\bq_1)\left(K^{(1)}_{\delta_g}(\bq_2) + f\mu_2^2 K_\theta^{(1)} (\bq_2)\right) + {\rm cyclic}\right]\,,\label{Ztwo}\\
Z^{(3)}(\bq_1,\bq_2,\bq_3) = \ & K^{(3)}_{\delta_g}(\bq_1,\bq_2,\bq_3) + f\mu_k^2 K^{(3)}_{\theta}(\bq_1,\bq_2,\bq_3)\nonumber\\
& + \mu_{k} k f \Big\{ \frac{\mu_1}{q_1} K^{(1)}_{\theta}(\bq_1) \left[K^{(2)}_{\delta_g}(\bq_2,\bq_3) + f\mu_{23}^2 K^{(2)}_{\theta}(\bq_2,\bq_3)\right]\nonumber\\
&+ \frac{\mu_{23}}{q_{23}}K^{(3)}_{\theta}(\bq_2,\bq_3) \left[K^{(1)}_{\delta_g}(\bq_1) + f\mu_1^2 K^{(1)}_{\theta}(\bq_1)\right] + {\rm cyclic}\Big\}\nonumber\\
&+ \mu_k^2 k^2 f^2\Big\{ \frac{\mu_2}{q_2}\frac{\mu_3}{q_3} K^{(1)}_{\theta}(\bq_2)K^{(1)}_{\theta}(\bq_3)\left[K^{(1)}_{\delta_g}(\bq_1) + f\mu_1^2 K^{(1)}_{\theta}(\bq_1)\right]+{\rm cyclic}\Big\}\,,
\label{rsdZ}
\end{align}
where we have suppressed the dependence on $a$ to simplify the notation, $\bk$ is the sum over internal momenta as in~\eqref{expansion_z2}, $\mu_k \equiv \hat \bk\cdot \hat {\bf z}$, $\mu_i \equiv \hat \bq_i\cdot \hat {\bf z}$, and $\mu_{ij} \equiv \hat \bq_{ij}\cdot \hat {\bf z}$. A hat denotes the unit vector, e.g.~$\hat \bk \equiv \bk/|\bk| $.

\subsection{Galaxy power spectrum in redshift space}
\label{bootPS}

The galaxy power spectrum in redshift space is defined by
\be
\label{genPS}
 \langle\delta_{g,s}(\bk;\eta)\delta_{g,s}(\bk';\eta)\rangle = (2\pi)^3\delta_D(\bk+\bk') P_{g, s}(\bk;\eta) \,.
\ee
Using the perturbative expansion in eq.~\eqref{expansion_z2}, it is possible to calculate the 1-loop power spectrum in perturbation theory as  the sum of three contributions,
\be
\label{P1loop}
P_{g,s}^{\rm 1-loop, PT}(\bk;\eta ) = P_{11}(\bk;\eta) + P_{22}(\bk;\eta) + P_{13}(\bk;\eta)\,.
\ee
The first term on the right-hand side is given by
\be
\label{P11}
P_{11}(\bk;\eta) = Z_1(\bk)^2 P_L(k; \eta)\,,
\ee
where $P_L(k; \eta)$ is the  linear power spectrum, defined by
\be
\label{linPS}
\langle\delta^{(1)}_\bk ( \eta)\delta^{(1)}_{\bk'} ( \eta)\rangle = (2\pi)^3 \delta_D(\bk + \bk') P_L(k;  \eta)\,.
\ee
Moreover,
\be
\label{P22}
P_{22}(\bk;\eta) = 2\int \frac{d^3 \bq}{(2\pi)^3}\left[Z_2(\bk - \bq,\bq)\right]^2 P_L(q;\eta) P_L(|\bk - \bq|;\eta)\,,
\ee
and
\be
\label{P13}
P_{13}(\bk;\eta)  = 6 Z_1(\bk)P_L(k;\eta)\int\frac{d^3\bq}{(2\pi)^3} Z_3(\bk, \bq, -\bq)P_L(q;\eta)\,.
\ee

The result  in eq.~\eqref{P1loop} needs to be corrected with  terms that account for the effects of the UV physics on the long modes, namely the counterterms  and the stochastic terms resulting from the stress-energy tensor in eq.~\eqref{finaleulereq} and from the bias and redshift-space expansions \cite{Senatore:2014vja,Perko:2016puo,Donath:2020abv},
\be
P_{g,s}^{\rm 1-loop}(\bk;\eta) = P_{g,s}^{\rm 1-loop, PT}(\bk;\eta) + P_{g,s}^{\rm CT}(\bk;\eta) + P_{g,s}^{\epsilon}(\bk;\eta)\;.
\label{fullPS}
\ee
At this order, the counterterms  in redshift space are
\be
P_{g,s}^{\rm CT}(\bk;\eta) = 2P_L(k;\eta)Z^{(1)}(\bk;\eta) \left(\frac{k}{k_{\rm M}}\right)^2 \left(  c_{\rm ct} + c_{r,1} \mu_k^2 + c_{r,2} \mu_k^4 \right)\,,
\label{Pct}
\ee
where $k_{\rm M}$ is the typical comoving scale of halos, and the stochastic contribution is given by
\be
P_{g,s}^{\epsilon}(\bk;\eta) = \frac{1}{\bar{n}_g}\left(c_{\epsilon,0} + \left(\frac{k}{k_{\rm M}}\right)^2c_{\epsilon,1} + f\mu_k^2\left(\frac{k}{k_{\rm M}}\right)^2c_{\epsilon,2}\right)\,,
\label{Peps}
\ee
where $\bar{n}_g$ is the mean number density of galaxies. 
From the anisotropic galaxy power spectrum, $P_{g,s}^{\rm 1-loop}(k, \mu_k)$, we can then calculate the multipoles,
\be
\label{multi}
P_{g,s}^{{\rm 1-loop,}(l)}(k,z) = \frac{2l+1}{2}\int^{1}_{-1}d\mu_k P_{g,s}^{\rm 1-loop}(k, \mu_k;z)\mathcal{P}_l(\mu_k)\;,
\ee
where $\mathcal{P}_l(\mu)$ are the Legendre polynomials of order $l$. In this work we will only consider the monopole and the quadrupole of the galaxy power spectrum. 

In addition, we perform the resummation of IR modes to account for the effect of long wavelength bulk motions on small scales \cite{Baldauf:2015xfa, Senatore:2017pbn}. Again, as for the bias expansion, for scale-indepedent models the IR resummation procedure is not modified in nDGP with respect to $\Lambda$CDM  \cite{DAmico:2021rdb}, and therefore we follow the scheme described in \cite{Lewandowski:2018ywf}, already implemented in the \code{PyBird} code.


\section{\code{PyBird} meets nDGP}
\label{sec:PB_nDGP}

In this section we introduce the code used, \code{PyBird}, and  specify the equations used to implement the nDGP model. Finally, in sec.~\ref{sub3.3}, we present the parameters used in the analysis and their priors. 

\subsection{The code}

\code{PyBird}~\cite{DAmico:2020kxu} is a code\footnote{See \href{https://github.com/pierrexyz/PyBird}{here} for the public GitHub repository.} based on Refs.~\cite{Perko:2016puo,DAmico:2019fhj}, written in Python, to compute the multipoles of the power spectrum of biased tracers in redshift space. The theoretical model is built on the EFTofLSS and a perturbative bias expansion scheme presented explicitly in~\cite{Donath:2020abv}.  
For a rapid evaluation, the  loop integrals (and the resummation integrals) are computed using the FFTLog method \cite{Simonovic:2017mhp}. 
See  \cite{Chudaykin:2020aoj, Chen:2020fxs,Chen:2020zjt} for other publicly available codes based on the EFTofLSS.

We have already discussed in sec.~\ref{sec:PTMG} the changes to the perturbation theory equations required by scale-independent modifications of gravity.  Fortunately, thanks to the modular nature of the code it is easy to implement  these extensions. In particular, the task is greatly facilitated by the fact that the perturbation theory kernels in the latest version of \code{PyBird} are computed taking into account their time dependence exactly, as opposed to the commonly used Einstein-de Sitter approximation.  One of the main modifications  consists in solving numerically the differential equation for the growth function and the  Green's functions in the nDGP case and replacing the ones computed in $\Lambda$CDM. We refer the reader to app.~\ref{app:IC_nDGP} for a discussion on how to choose the initial conditions of the linear solutions. 
The procedure, explained here for nDGP, can be easily extended to any modified gravity model with scale-independent linear growth:  Jordan Brans-Dicke~\cite{Brans:1961sx} and scalar-tensor theories with scalar fields of horizon-sized Compton wavelengths,  clustering quintessence~\cite{Sefusatti:2011cm, Anselmi:2011ef} (see~\cite{DAmico:2020tty} for the implementation of this model in \code{PyBird} and constraints on it with BOSS data in combination with BAO), the EFT of dark energy beyond linear order \cite{Cusin:2017wjg, Cusin:2017mzw,Bose:2018orj},
dark scattering models~\cite{Carrilho:2021hly}, k-mouflage theories~\cite{Babichev_2009,Brax_2014,Brax:2015pka,Benevento:2018xcu}, etc.

\subsection{The model}

In nDGP,\footnote{See \cite{Koyama:2005kd,Lombriser:2009xg,Xu:2013ega,Schmidt:2009sv} for a  treatment of cosmological perturbations of nDGP.} 
it is customary to introduce   the dimensionless  cosmological parameter $\Omega_{\rm rc} $ to measure the strength of modifications of gravity  in the effective 4d theory. This is  defined in eq.~\eqref{Orcdef}, where $r_c$ is the cross-over scale between the 5d and 4d cosmological behaviour, appearing in the  effective 4d Friedman equation as $H^2 +\frac{H}{r_c}= \frac{8 \pi G}{3} \sum_i \rho_i $ \cite{Deffayet:2000uy}, where $\rho_i$ are the background energy densities, including dark energy.

In this paper we are interested in constraining the effect of modified gravity on perturbations {\em only}. We will therefore assume that the background expansion is exactly the one of a flat-$\Lambda$CDM model, that is,
\be
\label{Hubble}
H (a) =H_0 \sqrt{\Omega_{\rm m} (a/a_0)^{-3} +  1-\Omega_{\rm m}  }  \;,
\ee
where $\Omega_{\rm m}$ is the present matter abundance. 
Such a behaviour can be realized by considering a dark energy component with fine-tuned dynamics \cite{Schmidt:2009sv}.
Modifications of gravity are therefore restricted to linear and higher-order perturbations, described by the single parameter $\Omega_{\rm rc}$.

Indeed, it can be shown (see e.g.~\cite{Bose:2018orj} and references therein for details) that the modified  Poisson equation in Fourier space is given by eq.~\eqref{newd2phi} with
\begin{align}
\label{mus}
\nu (a) &=   1+  \frac{1}{3 \beta(a)}   \;, \\
\label{mu2}
\nu_{2} (a) &= -\frac12  \left( \frac{H(a)}{H_0} \right)^2  \frac{1}{ \Omega_{\rm rc}} \left( \frac{1}{3 \beta(a)} \right)^3  \;,  \\
\label{mu22}
 \nu_{22}(a) & = 2  \left( \frac{H(a)}{H_0} \right)^4  \frac{1}{ \Omega_{\rm rc}^2} \left( \frac{1}{3 \beta(a)} \right)^5  \;
\end{align}
(as discussed above, we do not need to specify $\nu_{3}$),
where
\be
\beta(a) \equiv 1 + \frac{H(a)}{H_0} \frac{1}{\sqrt{\Omega_{\rm rc}}} \left( 1 + \frac{a H'(a)}{3H(a)} \right) \;.
\ee
One sees that GR is recovered by sending $H_0 r_c \to \infty$ or, equivalently, $\Omega_{\rm rc} \to 0$, which implies  $\beta \to \infty$ and consequently $\nu = 1$, $\nu_{2}= \nu_{22}=0$.

To solve the perturbation equations we need an expression for the time-dependent $\Omega_{{\rm m},a} (a)$ in terms of its value today, $\Omega_{{\rm m}} = \Omega_{{\rm m},a} (a=0)$. Since in this model $M$ is constant, from eq.~\eqref{Omdef} we find
\be
\Omega_{{\rm m},a} (a)= \Omega_{{\rm m}} \left( \frac{H_0}{H(a) } \right)^2 \left( \frac{a_0}{a}  \right)^3 \;.
\ee

\subsection{Parameters and priors}
\label{sub3.3}

Following \cite{DAmico:2019fhj}, we perform the analysis using the bias parameters
\be
\ctwo \equiv \frac{1}{\sqrt{2}} (\btwo+\bfour) \;, \qquad \cfour \equiv \frac{1}{\sqrt{2}} (\btwo- \bfour) \;,
\ee
instead of $\btwo$ and $\bfour$. Indeed, it was shown in that reference that 
$\btwo$ and $\bfour$ are highly degenerate in the data because they enter eqs.~\eqref{Kdg2} and~\eqref{Kdg3}  in such a way that their difference, $\cfour$, multiplies a term that is too small to be constrained by BOSS-like data. Here we set $\cfour=0$.
The parameters used in the analysis are 
\begin{equation}
\left\{\omega_{\rm b} , \text{ } \omega_{\rm cdm}, \text{ } h, \text{ }  A_{s} ,\text{ } n_s,\text{ }\log_{10} \Omega_{\rm rc},\text{ }\bone, \text{ } \ctwo, \text{} b_3, \text{} c_{\rm ct}, \text{} c_{\rm r,1}, \text{} c_{\rm \eps,0}, \text{} c_{\rm \eps,1}, \text{} c_{\rm \eps,2} \right\} \,.
\end{equation}
Since $\Omega_{\rm rc} \geq 0$, we decide to use $\log_{10} \Omega_{\rm rc}$ to scan many orders of magnitude. We then marginalize analytically on the remaining bias parameter, $\bthree$,  and on the   parameters appearing in eqs.~\eqref{Pct} and \eqref{Peps}, see below.

\begin{figure}[t]
\centering
    \includegraphics[width = 0.8\textwidth]{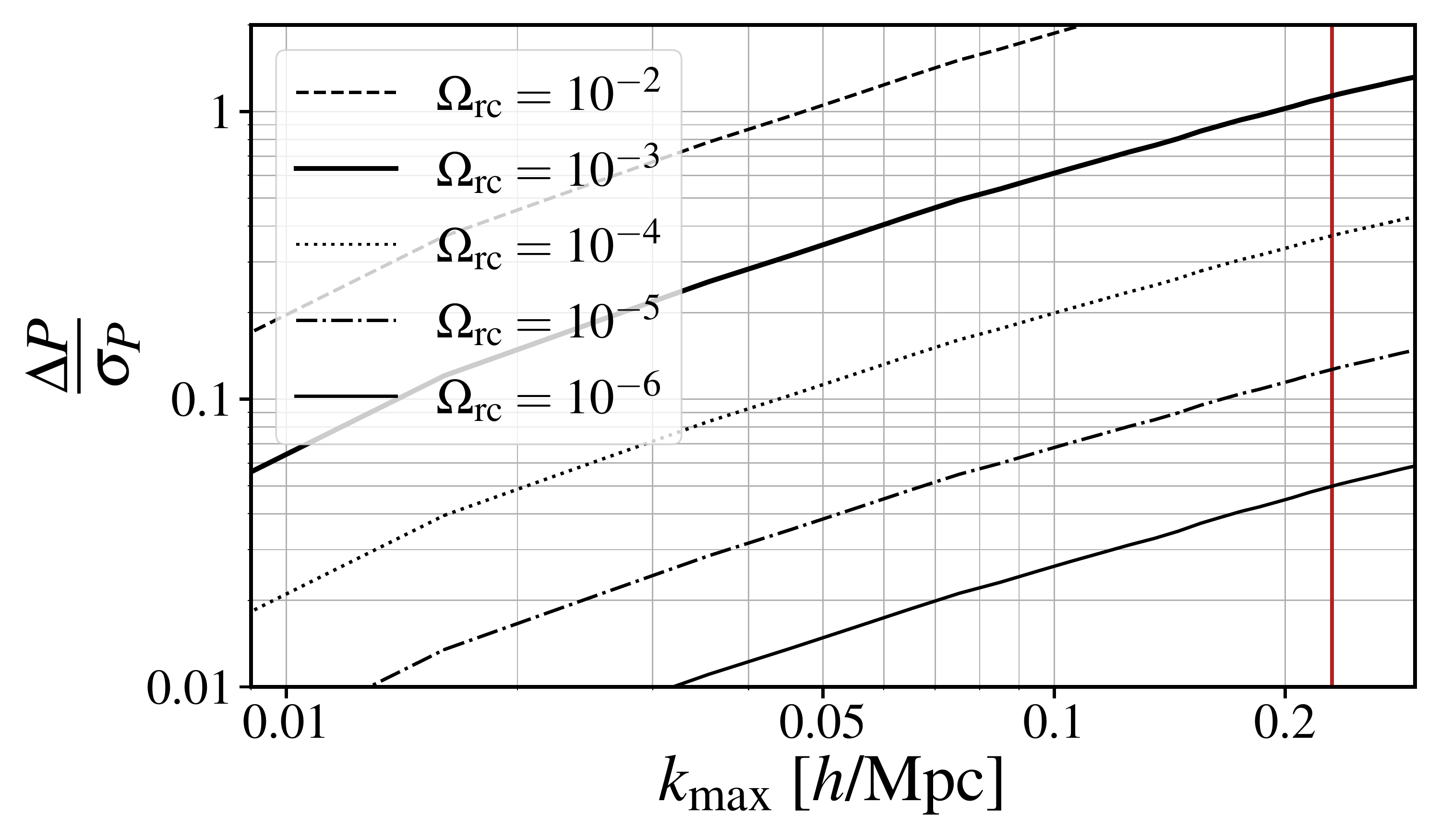}
    \caption{Ratio between  $\Delta P \equiv P_{g,s}^{\text{nDGP}, (0)}  -P_{g,s}^{\Lambda\text{CDM},(0)} $ and the  error $\sigma_P$ for the BOSS covariance at redshift $z = 0.61$, for different values of $\Omega_{\rm rc}$, and with bias parameters fixed to reproduce the monopole of the PT challenge simulations~\cite{Nishimichi:2020tvu}.   The red vertical line represents the $k_{\rm max}$ used in our analysis. Similar results are obtained at redshift $z = 0.38$.}
    \label{fig:OmrcSens}
\end{figure}
Let us now discuss the priors on these parameters. Using the BBN constraint of Ref.~\cite{Cooke:2017cwo}, we adopt a Gaussian prior on the absolute density of baryonic matter $\omega_{\rm b} = 0.02237 \pm 0.00036$.
We assume a flat prior for all the other parameters, except $A_s$ discussed below.
In the BOSS analysis, the priors  are
\be
\begin{split}
\omega_{\rm cdm}  \in \left[ 0.04,   0.25\right]
\;, \qquad h & \in \left[ 0.5,   1.0\right]
\;, \qquad n_{s} \in \left[ 0.5,   1.5\right] 
\;, \\ 
\log_{10}\Omega_{\rm rc}  \in \left[ -3,  2 \right]
\;, \qquad \bone & \in [0.8, 4] 
\;, \qquad \ctwo \in [-4, 4] 
\;. 
\end{split}
\ee
The lower and upper limits for the priors on $\bone$ and $\ctwo$ are motivated by previous  analysis, see~\cite{Nishimichi:2020tvu}. As can be seen in fig.~\ref{fig:boss_2},  the 2-$\sigma$ posterior distributions for these parameters are well inside the priors bounds, and therefore we conclude that these priors are very uninformative.
Moreover, we choose ${\rm min}\left[{\rm log}_{10}\Omega_{\rm rc}\right] = -3$ because our analysis is insensitive to lower values. Indeed, fig.~\ref{fig:OmrcSens} shows $\Delta P/\sigma_P$, where  $\Delta P \equiv   P_{g,s}^{\text{nDGP}, (0)}  -P_{g,s}^{\Lambda\text{CDM},(0)} $,  and $\sigma_P$ is the error on $P_{g,s}^{(0)}$ for the  BOSS volumes and galaxy densities,  as a function of $k_{\rm max}$, for different values of $\Omega_{\rm rc}$. For $\Omega_{\rm rc}\lesssim10^{-3}$, $\Delta P/\sigma_P \lesssim 1$, i.e.~the effect of modified gravity on the monopole is smaller than the error.   Adding the quadrupole does not change this conclusion.
 We discuss the implications of a different choice of ${\rm min}\left[{\rm log}_{10}\Omega_{\rm rc}\right]$  in sec.~\ref{sec:BOSS}.

Assuming that the typical comoving scale of halos appearing in eqs.~\eqref{Pct} and \eqref{Peps} is $k_{\rm M} = 0.7 \, h/\textrm{Mpc}$, we set gaussian priors centered on $0$ on the EFT parameters, with widths given by 
\be
\begin{split}
\sigma({\bthree}) &= 2   \;, \qquad \sigma({c_{\rm ct}})  = 2  \;, \qquad \sigma({c_{ r,1}}) = 8 \;,  \\ \sigma({c_{  \epsilon,0}}) &= 2 \;, \qquad \sigma({c_{  \epsilon,1}})  = 2  \;, \qquad \sigma({c_{  \epsilon,2}}) = 2\;,
\end{split}
\ee
and we analytically marginalize over them, as described  in~\cite{DAmico:2020kxu}. Since we do not compute the hexadecapole, $c_{r,1}$   and $c_{r,2}$ are completely degenerate: we absorb the latter into the former and enlarge the prior width  to $8$. We choose a galaxy number density $\bar{n}_g = 3 \times 10^{-4} (\textrm{Mpc}/h)^3$ for the analysis of the synthetic data and the PT Challenge simulations at  redshifts $z = 0.38$ and $z = 0.61$ respectively, and we use $\bar{n}_g = 4 \times 10^{-4} (\textrm{Mpc}/h)^3$ and $\bar{n}_g = 4.5 \times 10^{-4} (\textrm{Mpc}/h)^3$ for the analysis of the BOSS catalogues~\cite{BOSS:2016psr}, at  redshifts $z = 0.38$ and $z = 0.57$ respectively. We apply the Alcock-Paczynski effect to all the analyses performed in this work.

Finally, for the primordial amplitude of scalar fluctuations, $A_{s}$, we present  two cases:
\begin{enumerate}
    \item Assuming a flat prior in the range $A_s \in \left[ 0.1,  \text{ }5.0  \right] \times 10^{-9} $;
 \item Fixing it to the Planck central value \cite{Aghanim:2018eyx}: $ A_s = 2.0989 \times 10^{-9} $.
\end{enumerate}
For the second case, given the smallness of Planck's error on $A_s$, $1.4 \%$, fixing it or assuming a gaussian prior with $1\, \sigma$ width essentially  gives the same results, as we have explicitly checked.  

We run MCMC's   based on the Metropolis-Hastings sampler as implemented in \code{MontePython 3}~\cite{Audren:2012wb,Brinckmann:2018cvx}, with the theory model evaluated using \code{CLASS}~\cite{Blas:2011rf} and \code{PyBird}.
We declare convergence of the chains when the Gelman-Rubin $R-1$ value~\cite{Gelman:1992zz} is lower than $0.02$.
We used \code{GetDist}~\cite{Lewis:2019xzd} package  to create the plots and  calculate the summary statistics.

\section{Testing the pipeline}
\label{sec:pipeline}

Before applying our pipeline to the real BOSS data, we have performed several tests on synthetic data and simulations. The aim is twofold: exploring projection effects due to strong degeneracy between parameters, such as for instance between $A_s$ and $\Omega_{\rm rc}$ as explained in the next subsection, and assessing the maximal wavenumber that we can use in our analysis, $k_{\rm max}$.
Indeed, in sec.~\ref{synthdata} we test our pipeline on synthetic data and in sec.~\ref{sec: sims} we analyse the PT Challenge simulations.  
Finally, in sec.~\ref{sub4.4} we discuss projection effects in more details.

\subsection{Degeneracy between $A_s$ and $\Omega_{\rm rc}$}
\label{sub4.1}

\begin{figure}[t]
\centering
    \includegraphics[width = 0.8\textwidth]{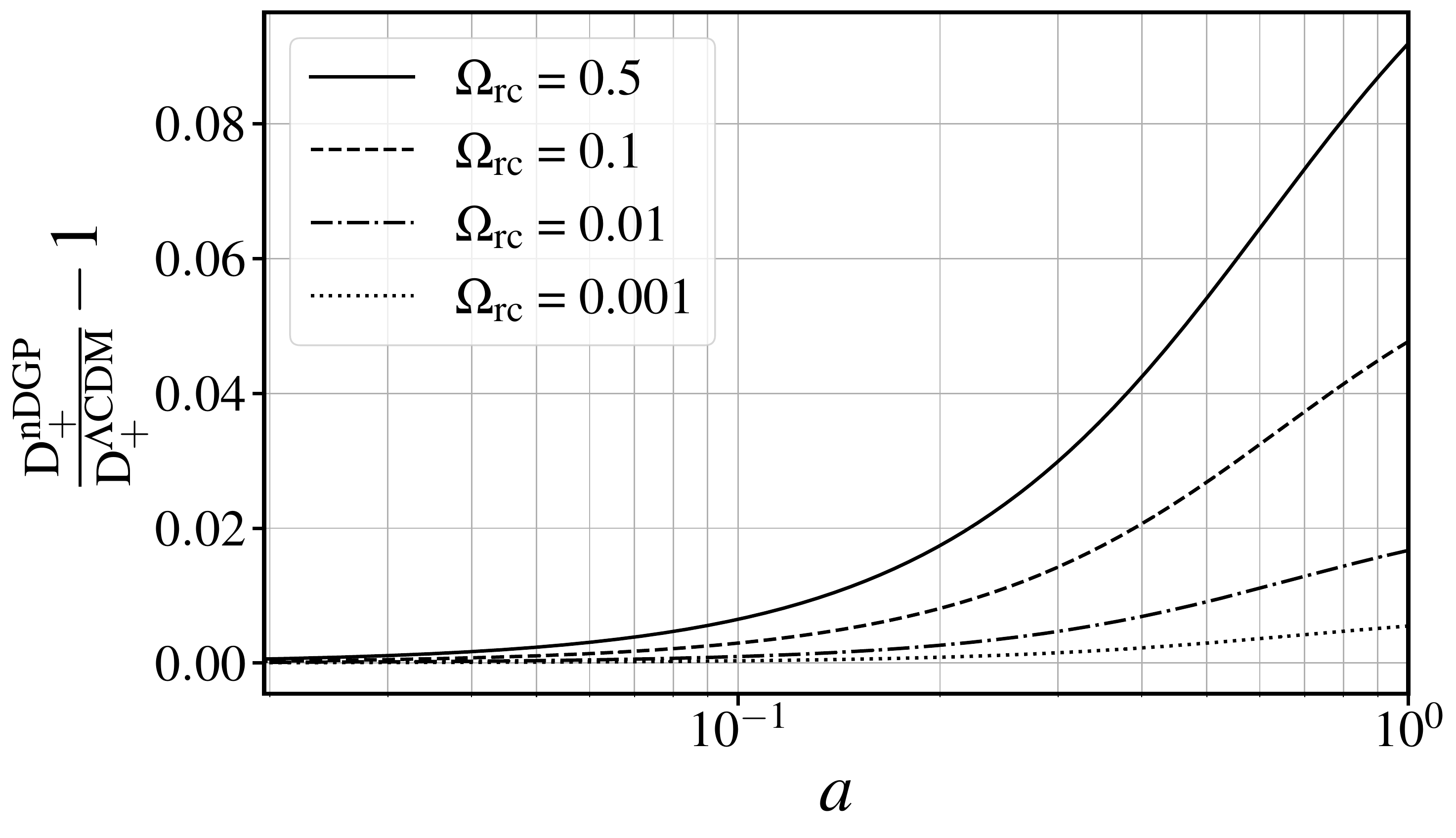}
    \caption{Growth function for nDGP with various values of $\Omega_{\rm rc}$ as a function of the scale factor $a$, rescaled to the $\Lambda$CDM one, with $\Omega_{\rm m}$ fixed to Planck's best-fit value. }
    \label{fig:dcompnDGP}
\end{figure}
We expect the nDGP parameter $\Omega_{\rm rc}$ to be degenerate with the amplitude of primordial fluctuations $A_s$. Indeed, the former enters the  modified linear growth equation, eq.~\re{growth}, via the function $\nu (a)$ introduced in \re{newd2phi}. In fig.~\ref{fig:dcompnDGP} we show the effect of modified gravity on the  linear growth function for different values of  $\Omega_{\rm rc}$.

\begin{figure}[ht]
    \centering
    \includegraphics[width = 0.8\textwidth]{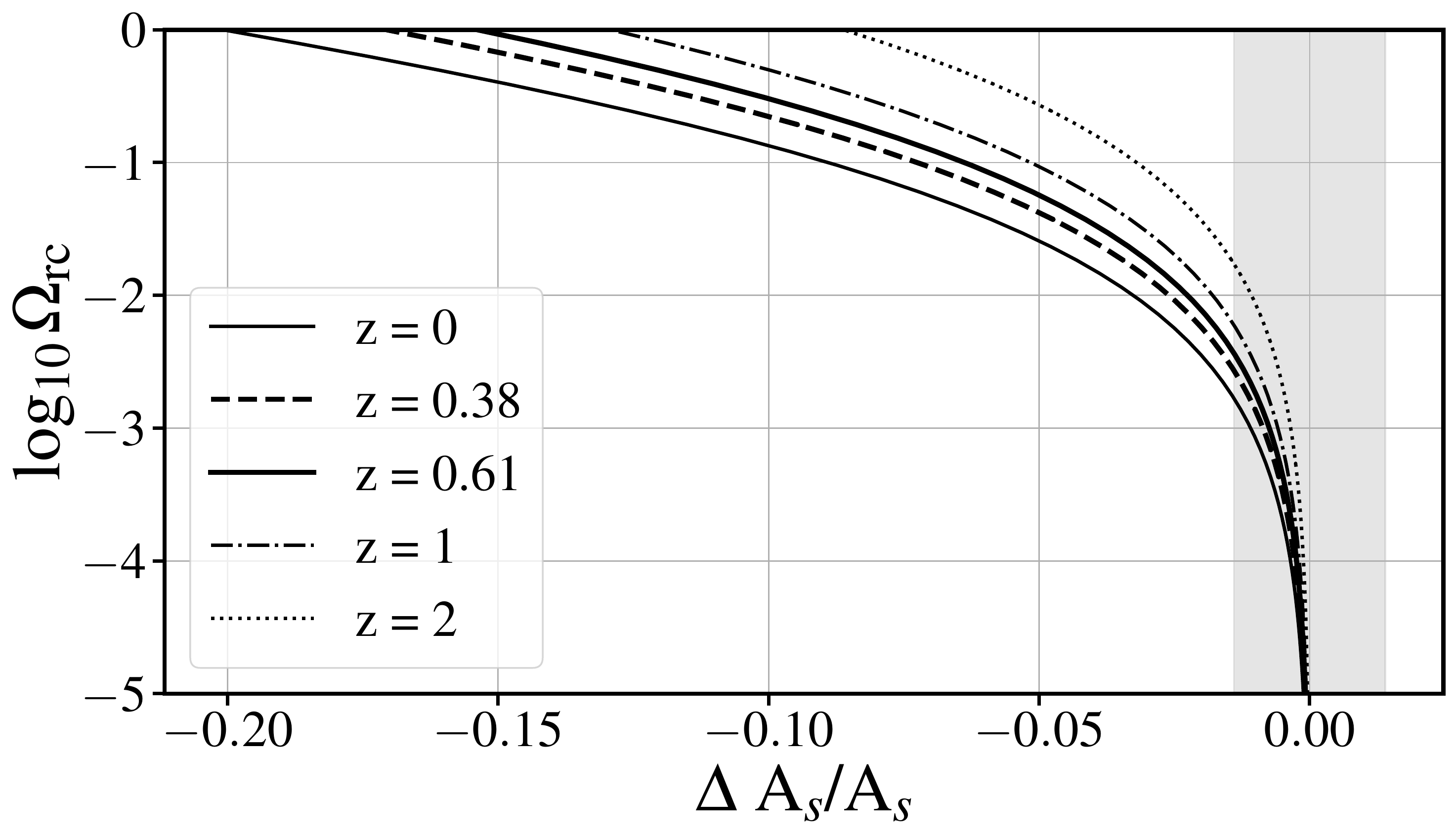}
    \caption{Curves of constant $D_+^2(z) A_s$ in the $\Delta A_s/A_s$-$\Omega_{\rm rc}$ plane, where $\Delta A_s/A_s \equiv \left(A_s^{\rm nDGP}-A_s^{\rm  \Lambda CDM} \right)/A_s^{\rm  \Lambda CDM}$ defines the shift in $A_s$ required to compensate  the change in $D_+^2(z)$, for different redshifts. The two thick lines represent the central redshifts of the PT simulations data samples considered in our analysis, which are very close to the BOSS catalogs analyzed in this work. The shaded region represents the Planck $\pm 1\sigma$ error. }
    
    \label{fig:degeneracy}
\end{figure}
Since the time dependence of the linear density power spectrum   is given by $P(k;z)\propto D_+^2(z) A_s$, the effect of $\Omega_{\rm rc}$ on the linear growth function is exactly degenerate with the primordial amplitude $A_s$.  To show the degeneracy as a function of the redshift, in fig.~\ref{fig:degeneracy} we plot curves of constant $D_+^2(z) A_s$ in the $\Delta A_s/A_s$-$\log_{10} \Omega_{\rm rc}$ plane, where we have defined the shift in $A_s$ required to compensate  the change in $D_+^2(z)$ as
\beq
\frac{\Delta A_s}{A_s} \equiv \frac{A_s^{\rm nDGP}-A_s^{\rm  \Lambda CDM}}{A_s^{\rm \Lambda CDM}}\,.
\eeq

Combining data at different redshifts can, in principle, break this degeneracy but this does not work if $\Omega_{\rm rc}$ is too small or if the redshift bins are too close. In fig.~\ref{fig:degeneracy} we also show the curves for $z = 1$ and $z= 2$.

\subsection{Null test on synthetic data}
\label{synthdata}

\begin{table}[t]
\centering
\setlength{\tabcolsep}{0.5em} 
\renewcommand{\arraystretch}{1.2}
\begin{tabular}{|l||c|c|c|c|c|} 
 \hline 
  & Synthetic data & PT-Challenge sims & BOSS data \\ \hline \hline 
Low-$z$ sample &   $ z=0.38$   & $z=0.38$   & $z=0.32$   \\ 
 &    $k_{\rm max} = 0.20 h \, \text{Mpc}^{-1} $  &  $k_{\rm max} = 0.20 h\, \text{Mpc}^{-1}$ &  $k_{\rm max} = 0.20 h \, \text{Mpc}^{-1}$  \\ \hline
High-$z$ sample  & $z=0.61$ & $z=0.61$ & $z=0.57$\\  
  & $k_{\rm max} = 0.23 h  \,\text{Mpc}^{-1}$  &  $k_{\rm max} = 0.23 h  \,\text{Mpc}^{-1}$  &  $k_{\rm max} = 0.23 h  \,\text{Mpc}^{-1}$ \\  \hline
\end{tabular}
\caption{Summary of the data samples considered for our analyses, with their respective redshifts and $k_{\rm max}$ used.}
\label{tab:data}
\end{table}

Using the EFTofLSS model described in sec.~\ref{sec:PTMG},  we have created synthetic data, with known cosmology and bias parameters, for  $\Lambda$CDM (i.e.~$\Omega_{\rm rc}=0$), and we have tested our pipeline on these data. 
The goal of this preliminary analysis is  to verify if  the marginalized posterior  probability are affected by strong projection effects, such as the one discussed above. See also e.g.~\cite{Gomez-Valent:2022hkb,DAmico:2022osl} for a discussion on this topic. 

We fixed the  cosmological parameters\footnote{Following the philosophy of \cite{Nishimichi:2020tvu}, we do not disclose their values here.} of the synthetic data to be the same as those of the PT Challenge mocks, while the bias and EFT parameters were chosen using the best-fit values obtained by fitting the mocks.
We generated samples at the two redshifts of the simulations dataset analyzed below, $z =0.38$ and $z=0.61$.
We have analyzed the monopole and quadrupole of the synthetic galaxy power spectrum for both redshift samples and for the combination of the two.
We have used a Gaussian covariance provided in  the PT-Challenge website\footnote{\url{https://www2.yukawa.kyoto-u.ac.jp/~takahiro.nishimichi/data/PTchallenge}}, rescaled in order to match the volume and number densities of the BOSS catalogues~\cite{BOSS:2016psr}.
We have fixed the maximum wavenumber included in the analysis to $k_{\rm max} = 0.23 \, h$/Mpc for $z = 0.61$ and $k_{\rm max} = 0.20 \, h$/Mpc for $z = 0.38$; for a summary, see tab.~\ref{tab:data}. As explained in sec.~\ref{sub4.4}, when analyzing the simulations we find that the theoretical systematic error on cosmological parameters is negligible.

\begin{figure}
\begin{center}
\includegraphics[width=0.79\textwidth]{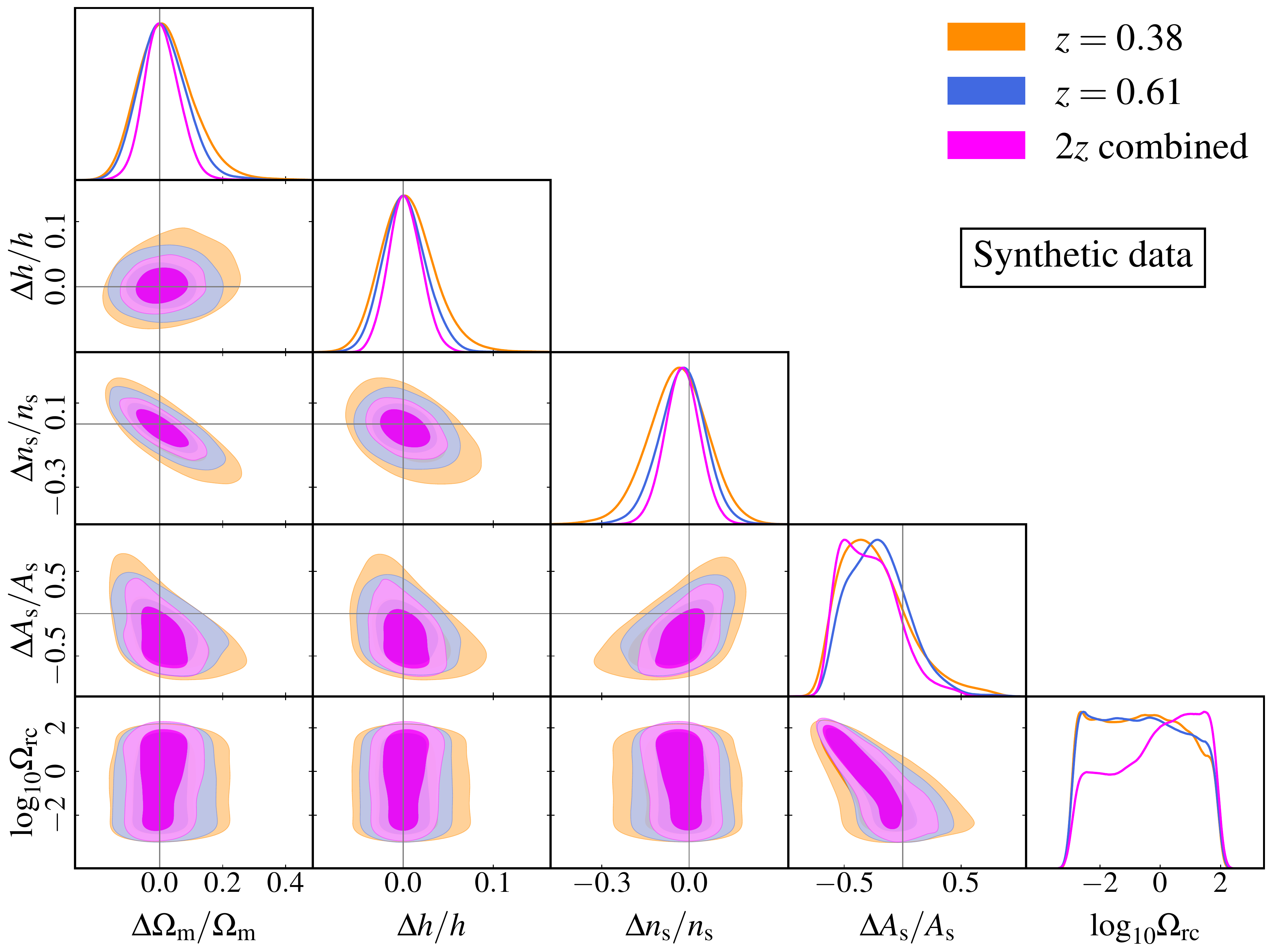}

\vspace{0.7cm}

\includegraphics[width=0.79\textwidth]{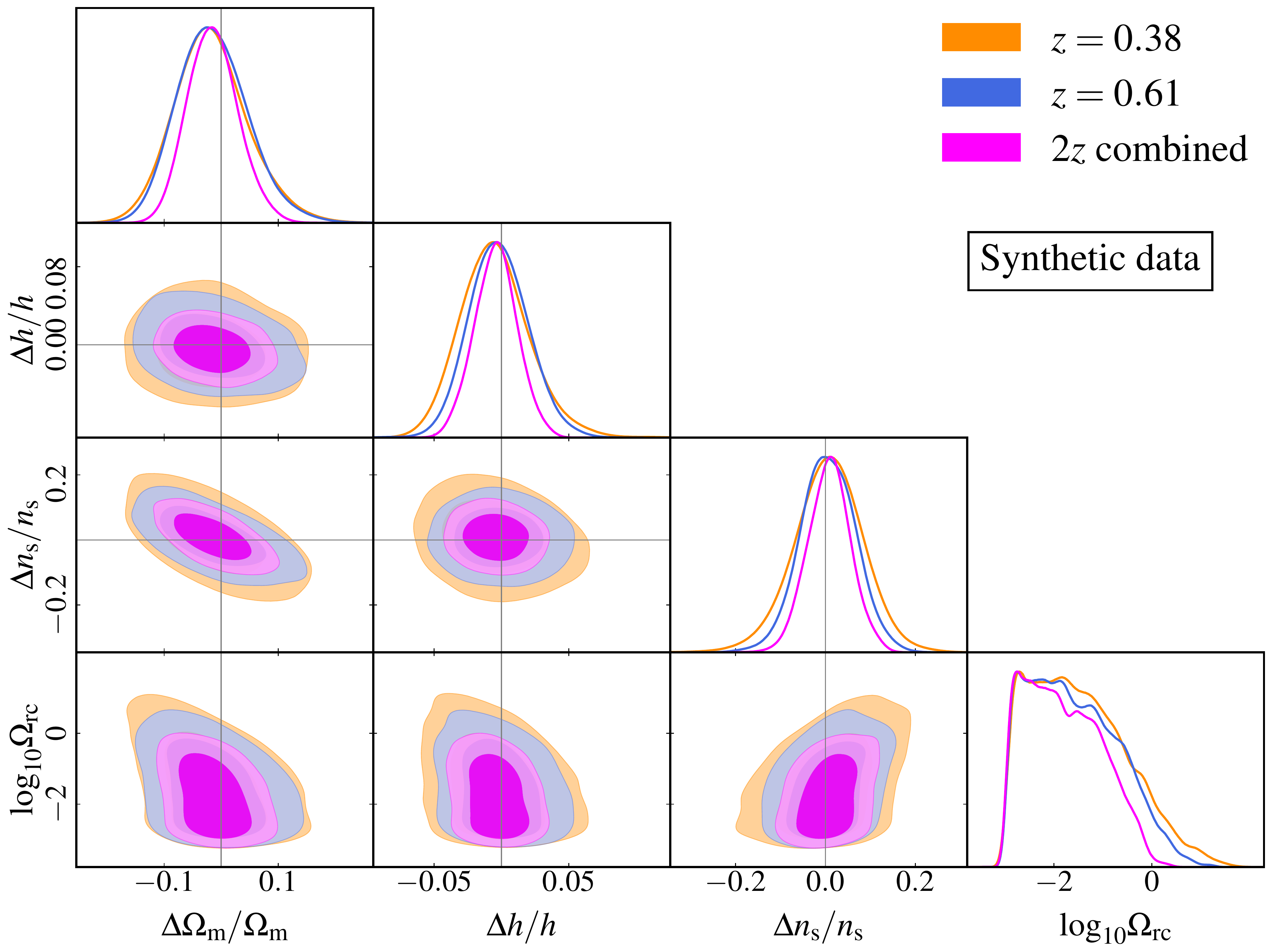}
\caption{Marginalized posteriors for the cosmological parameters from $\Lambda$CDM-synthetic data, with flat-$A_s$ priors (top panel) and fixed $A_s$ (bottom panel), with two different sky-cuts and their combination. The covariances used have been rescaled to match that of BOSS data.}
\label{fig:syntheticdata}
\end{center}
\end{figure}
The results  are shown in fig.~\ref{fig:syntheticdata}, with a flat prior on $A_s$ (top) and with fixed $A_s$ (bottom). 
The expected degeneracy between $\Omega_{\rm rc}$ and $A_{s}$ is visible  (top panel) for $\Omega_{\rm rc}\gtrsim 10^{-2}$, which results in a biased determination of $A_s$ and a spurious peak in the $\Omega_{\rm rc}$ posterior, induced by projection effects. As anticipated, the shape of the 2d posterior in the $\log_{10}\Omega_{\rm rc}$-$A_s$ plane is only mildly redshift dependent; therefore, combining the two relatively close redshift samples does not help in reducing the degeneracy.
Fixing $A_s$ to Planck's value (botton panel) clearly   reduces the degeneracy and removes projection effects. Moreover, increasing $\Omega_{\rm rc}$ while keeping $A_s$ fixed lifts the amplitude of the linear power spectrum, and therefore the relative weight of the 1-loop effects with respect to the linear ones, which is instrumental in lifting degeneracies among cosmological and bias parameters. Furthermore, it is now possible to observe that there is also a moderate degeneracy between $\log_{10}\Omega_{\rm rc}$, $\textit{n}_{s}$ and $\Omega_m$ that it is less relevant in the case in which $A_s$ is varied.


\begin{figure}
\centering
\includegraphics[width=0.79\textwidth]{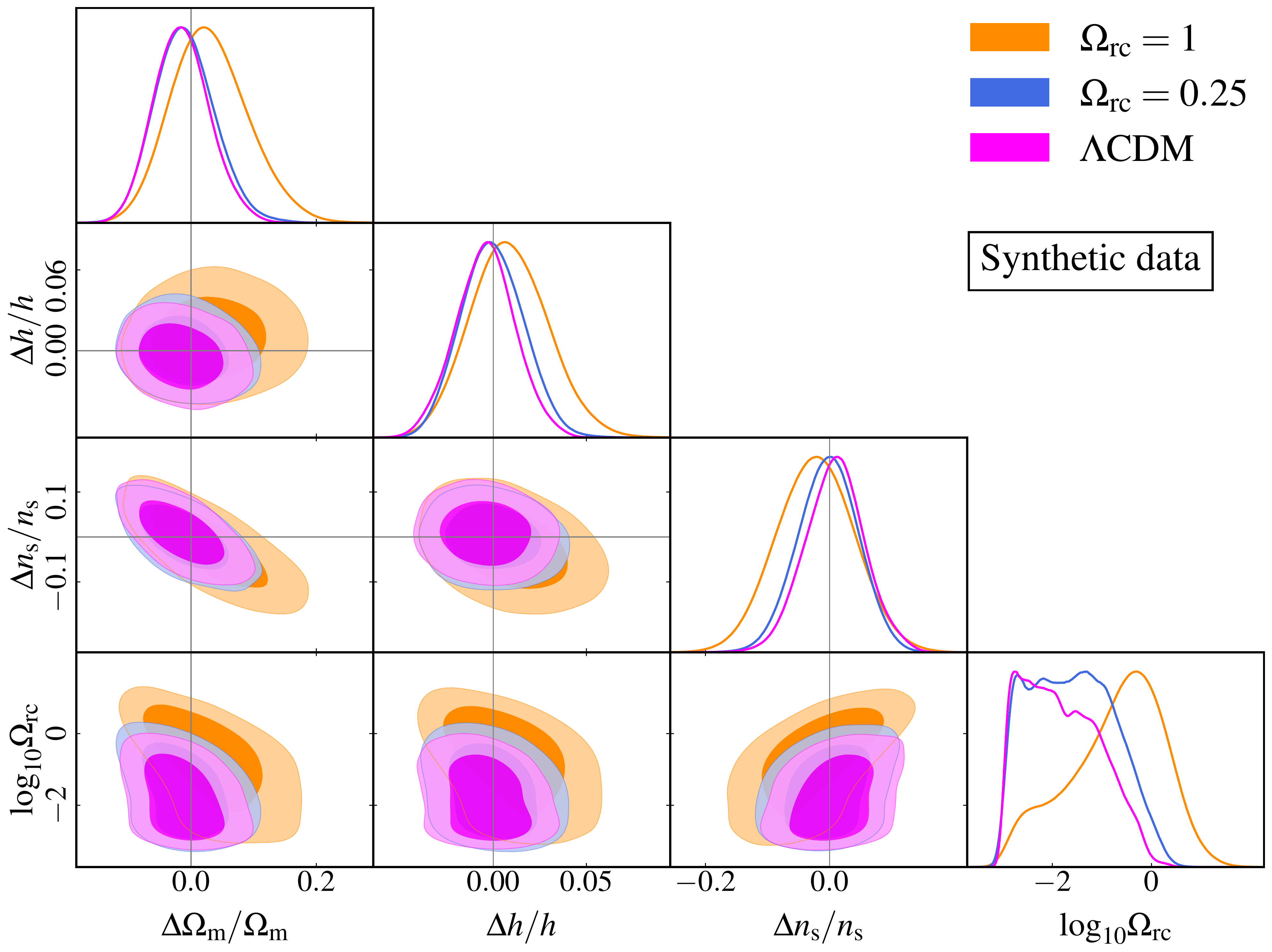}
\caption{Marginalized posteriors  for the cosmological parameters for nDGP with $\Omega_{\rm rc}=0.25$ and $\Omega_{\rm rc}=1$, compared to $\Lambda$CDM, for the combined sky-cuts and with $A_s$ fixed to Planck's value.}
\label{fig:triangle_plots_detection_BOSSCov_2z_As_fix_OMrc_vs_LCDM_comparison} 
\end{figure}
Additionally, we have performed an analysis of synthetic data in nDGP combining two redshift samples. To observe a clear effect of nDGP we have considered large values of $\Omega_{\rm rc}$, $\Omega_{\rm rc} = 1$ and $\Omega_{\rm rc} = 0.25$, leaving the other parameters as above and fixing $A_s$ to the Planck central value. In fig.~\ref{fig:triangle_plots_detection_BOSSCov_2z_As_fix_OMrc_vs_LCDM_comparison} we compare the marginalized posteriors of these two cases with the $\Lambda$CDM case.
A  clear detection is present only for values bigger than $\Omega_{\rm rc} \gtrsim 1$.\footnote{With \textit{detection} here we mean the following: we subtract the 2-$\sigma$ error to the   measured best fit value and we compare it with the lower bound of the prior used for the analysis. } 

We also note that for the cosmology with large $\Omega_{\rm rc}$ the uncertainties of the other cosmological parameters are larger due to the different degeneracies with the other cosmological parameters around the central value.

\subsection{Tests on simulations} 
\label{sec: sims}

\begin{figure}
\begin{center}
\includegraphics[width=0.79\textwidth]{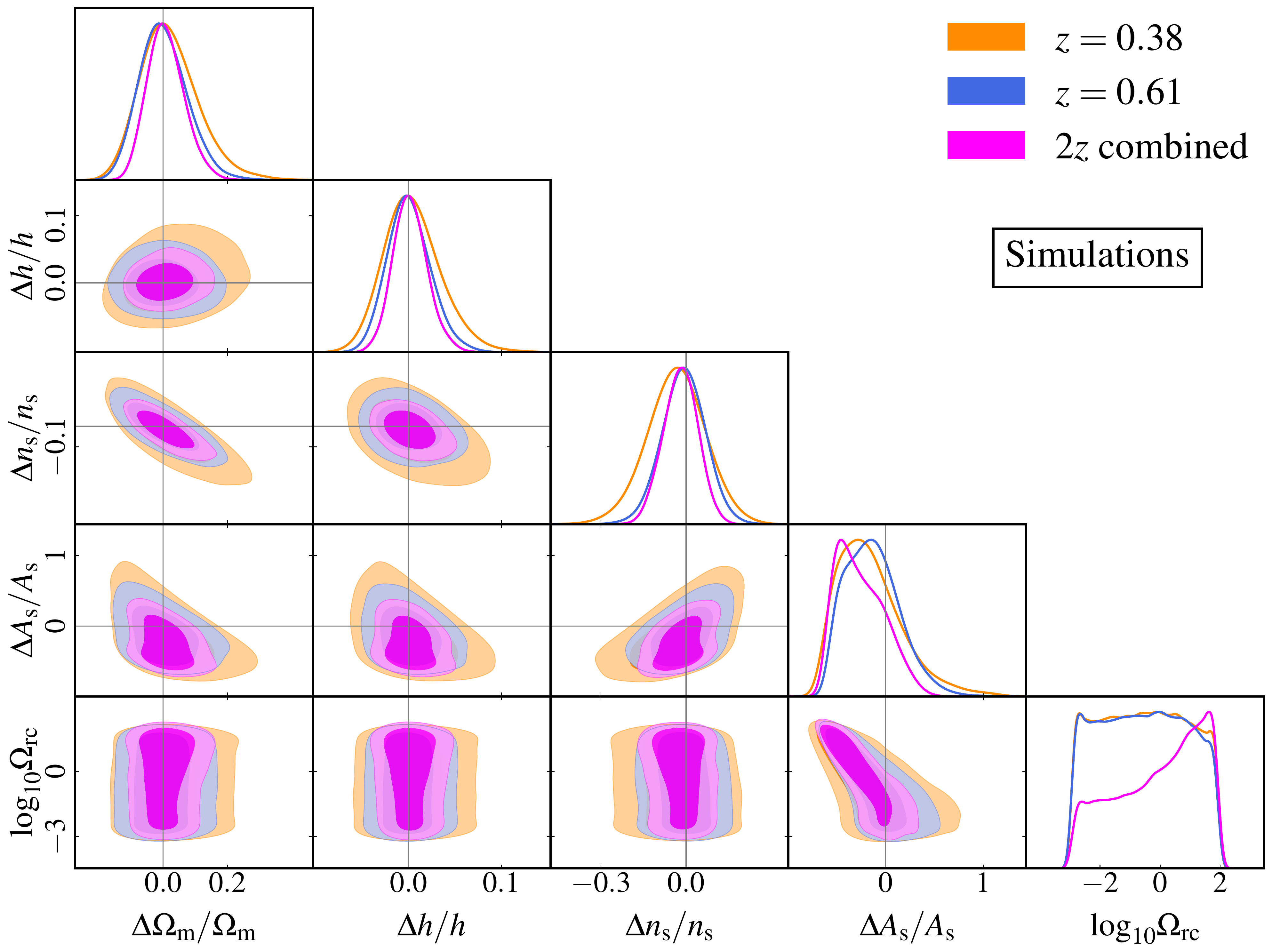}

\vspace{0.7cm}

\includegraphics[width=0.79\textwidth]{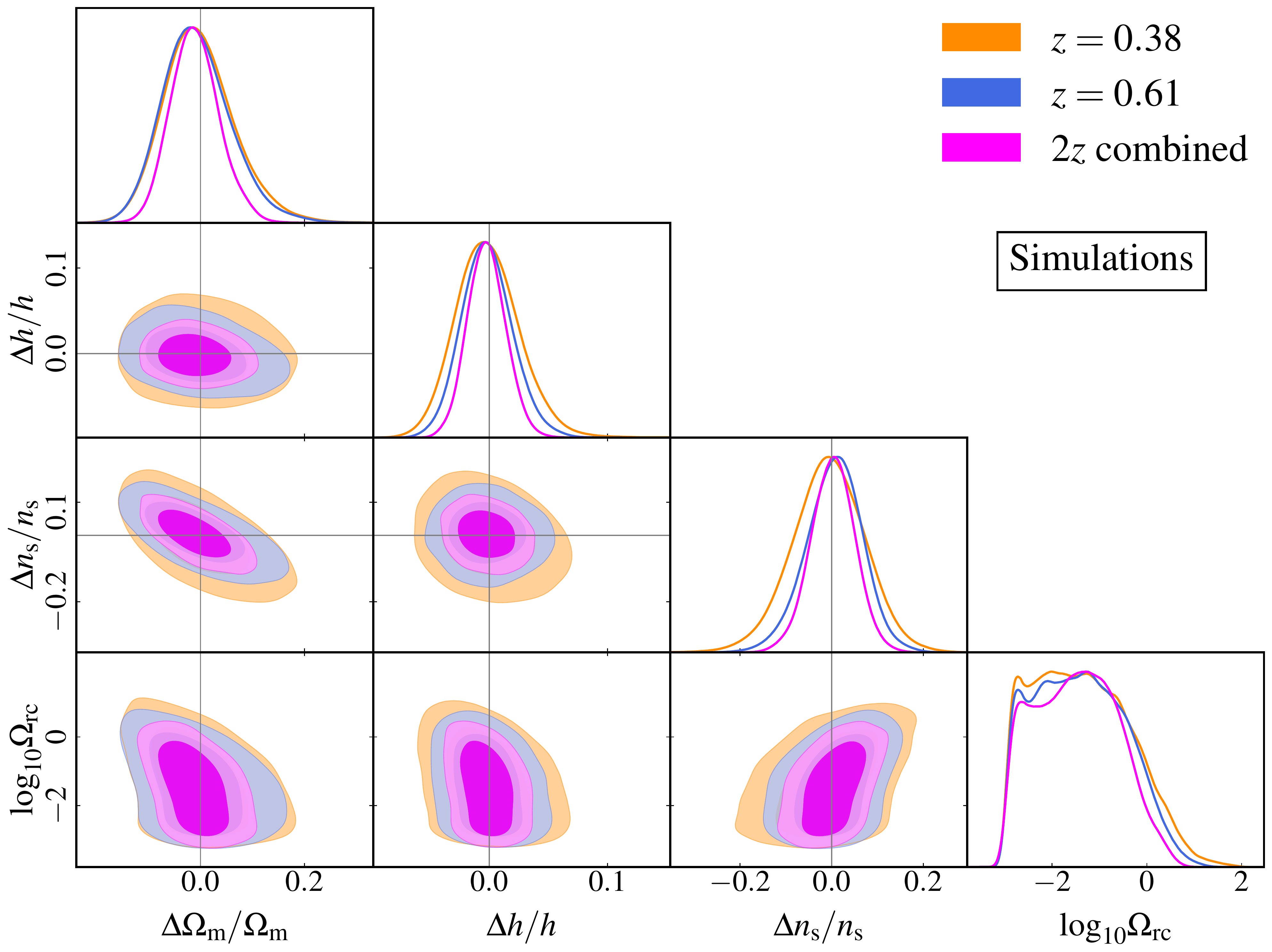}
\caption{Marginalized posteriors for the cosmological parameters from the PT Challenge simulations, with flat-$A_s$ priors (top panel) and fixed $A_s$ (bottom panel), with two different sky-cuts and their combination. The covariances used have been rescaled to match those of BOSS data.}
\label{fig:simulations}
\end{center}
\end{figure}
We tested our pipeline on the PT Challenge simulations \cite{Nishimichi:2020tvu}, which are ten realizations in periodic comoving boxes of side length of $3840 \text{ Mpc}/h$ with $3072^3$ particles each. The total volume, $566\, ({\rm Gpc}/h)^3$, is about a hundred times that of the BOSS catalogues. A flat $\Lambda$CDM cosmology is assumed, with $\Omega_{\nu}$ set to zero. Dark matter halos are  identified with the Rockstar halo finder~\cite{2013ApJ...762..109B} and then populated with mock galaxies matched to reproduce the observed clustering properties of the BOSS samples. Further details can be found in~\cite{Nishimichi:2020tvu}.
We analyzed these simulations with $k_{\rm max} = 0.23 \, h$/Mpc for $z = 0.61$ and $k_{\rm max} = 0.20 \, h$/Mpc for $z = 0.38$, as already described in the previous section.

The results are shown in fig.~\ref{fig:simulations} for two cases: varying $A_s$ with a flat prior (top) and for $A_s$ fixed to the Planck central value (bottom).
In the top panel we observe, again, the degeneracy between $\Omega_{\rm rc}$ and $A_{s}$ and the presence of a peak in the $\log_{10}\Omega_{\rm rc}$ posterior, which indicates the presence of projection effects. These effects disappear when fixing $A_s$ to Planck's value (bottom panel). As for synthetic data, 
a moderate degeneracy between $\log_{10}\Omega_{\rm rc}$, $n_{s}$ and $\Omega_m$ emerges when $A_s$ is fixed.

\subsection{Projection effects and theoretical errors}
\label{sub4.4}

\begin{table}[t]
\centering
\setlength{\tabcolsep}{0.5em} 
\renewcommand{\arraystretch}{1.2}
\begin{tabular}{|l||c|c|c|c|c|} 
 \hline 
$x $ & $ \Omega_{\rm m}$ & $h$ & $n_s$ \\ \hline  
$\Delta_{\rm synth}(x)/\sigma_{\rm synth}(x) $ &  $-0.2333$ & $-0.0903$ & $0.0938$ \\ 
$\Delta_{\rm sims}(x)/\sigma_{\rm sims}(x)$ & $-0.1053$ & $-0.0826$ & $0.0901$\\  
$\Delta_{\rm sims}(x)/\sigma_{\rm sims}(x) - \Delta_{\rm synth}(x)/\sigma_{\rm synth}(x)$ & $0.1280$ & $0.0077$ & $-0.0038$\\ \hline
\end{tabular}
\caption{Ratio $\Delta_{\rm data}(x)/\sigma_{\rm data}(x)$, where $\Delta(x)= \bar{x} - x_{\rm truth}$ is the difference between the marginalized mean and the true value, and $\sigma_{\rm data}$ is the data error,  measured on synthetic data (first row) and the PT Challenge simulations (second row), for $x = \Omega_{\rm m}$, $h$ and $n_s$. We consider only the high-$z$ sample and we fix $A_{s}$ to the Planck central value. We report also the difference between the two cases (third row).}
\label{tab:shifts_from_truth}
\end{table}

The above analysis shows a large difference between the mean (or the median) of the 1d marginalized posterior and the true value of $A_s$.
Following \cite{DAmico:2022osl}, to estimate the importance of other systematic biases due to projection effects, 
in tab.~\ref{tab:shifts_from_truth} we show the ratio between $\Delta(x)= \bar{x} - x_{\rm truth}$, i.e.~the difference between the marginalized mean value $\bar{x}$ of a parameter $x$ and its true value $x_{\rm truth}$, and the error associated to the data for the same parameter, $\sigma_{\rm data}(x)$.
These are computed for the synthetic data and for the  simulations analyses, considering only one sky-cut (i.e.~the high-$z$ case), for the case in which $A_{s}$ is fixed to the Planck value. We find that the deviations due to projection effects are negligible with respect the dataset errors, i.e., $\Delta_{\rm data}/ \sigma_{\rm data}(x)\lesssim 1/3$.

For the simulation analysis, there are two contributions to the deviations of the means from their true values: the projection effects and the theory error. 
To estimate the importance of the theory error, we subtract the shift measured with synthetic data (which are exempt from theoretical errors) divided by the error from the same ratio measured with simulations: this is safely negligible in our analysis, as shown in the table.

\section{Results}
\label{sec:BOSS}
\subsection{BOSS analysis}

In this section we apply our pipeline to the analysis of real data from the SDSS-III  BOSS \cite{Gil-Marin:2015sqa}.
In particular, we have analyzed the full-shape BOSS DR12 power spectrum measurements. 
The theory model is the same EFTofLSS used in the fit of synthetic data and simulations, with the same priors. 
We convolve the power spectrum multipoles with the survey window function measured with the technique outlined in~\cite{Beutler:2021eqq}, with a consistent normalization for the power spectrum estimator.\footnote{See \href{https://github.com/pierrexyz/fkpwin}{here} for the public GitHub repository for the evaluation of the window function.}

We analyze the monopole and quadrupole power spectra of the redshift cuts, $z_{\rm eff} = 0.57$ and $z_{\rm eff} = 0.32$, using both North Galactic Cap and South Galactic Cap sky-cuts for each redshift bin. 
We cut the power spectra at $k_{\rm max} = 0.23 \, h/\textrm{Mpc}$ for the higher redshift and at $k_{\rm max} = 0.20 h/\textrm{Mpc}$ for the lower one. When analyzing  BOSS data, we also include the effects of neutrinos in the  linear power spectrum. We use the Planck prescription of one massive neutrino species with $m_\mathrm{\nu} = 0.06 $ eV and two massless ones contributing to the number of effective relativistic species as $N_{\rm eff} = 2.0328$.

\begin{figure}
\begin{center}
\centering
\includegraphics[width=0.79\textwidth]{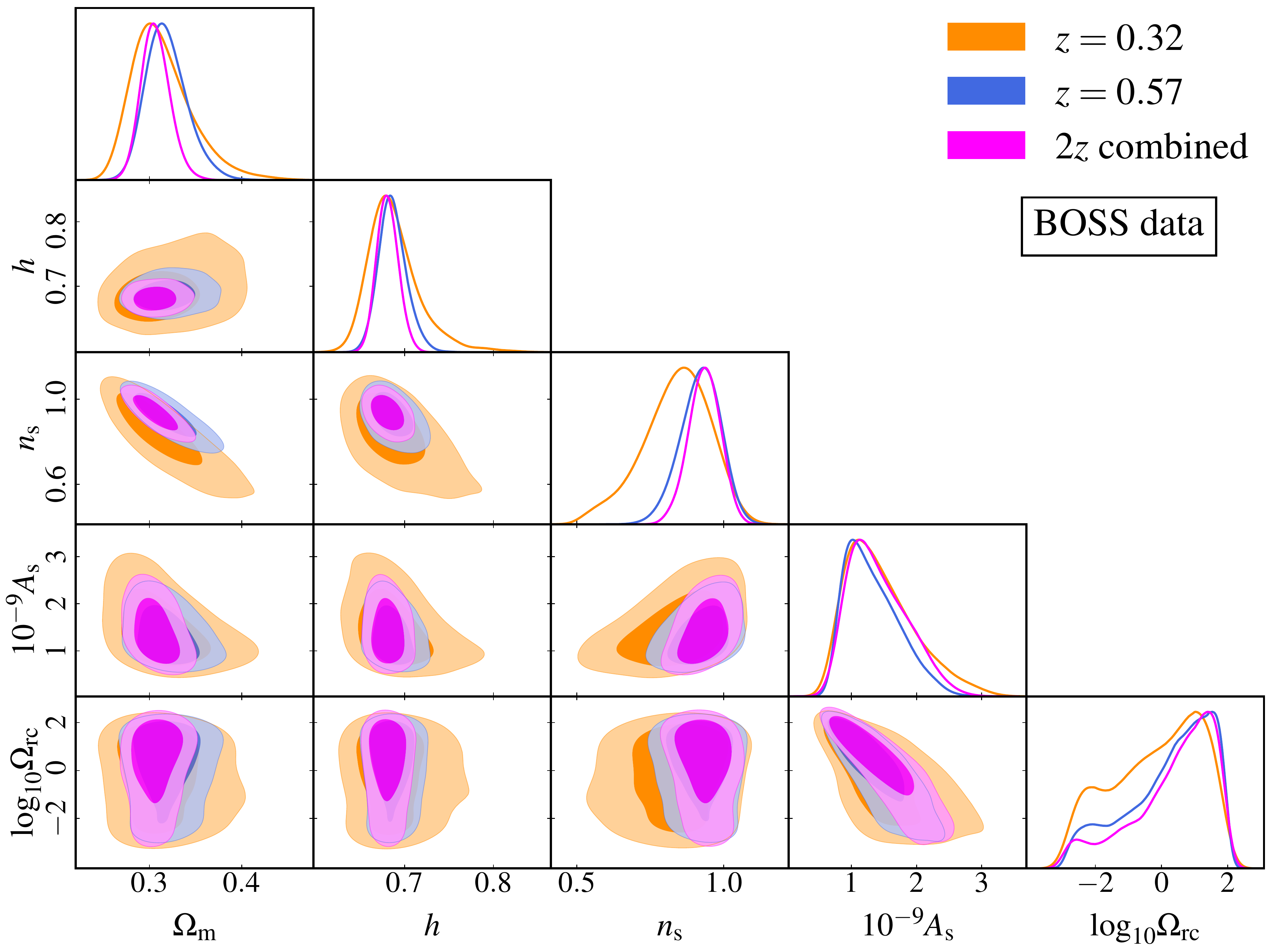}

\vspace{0.7cm}

\includegraphics[width=0.79\textwidth]{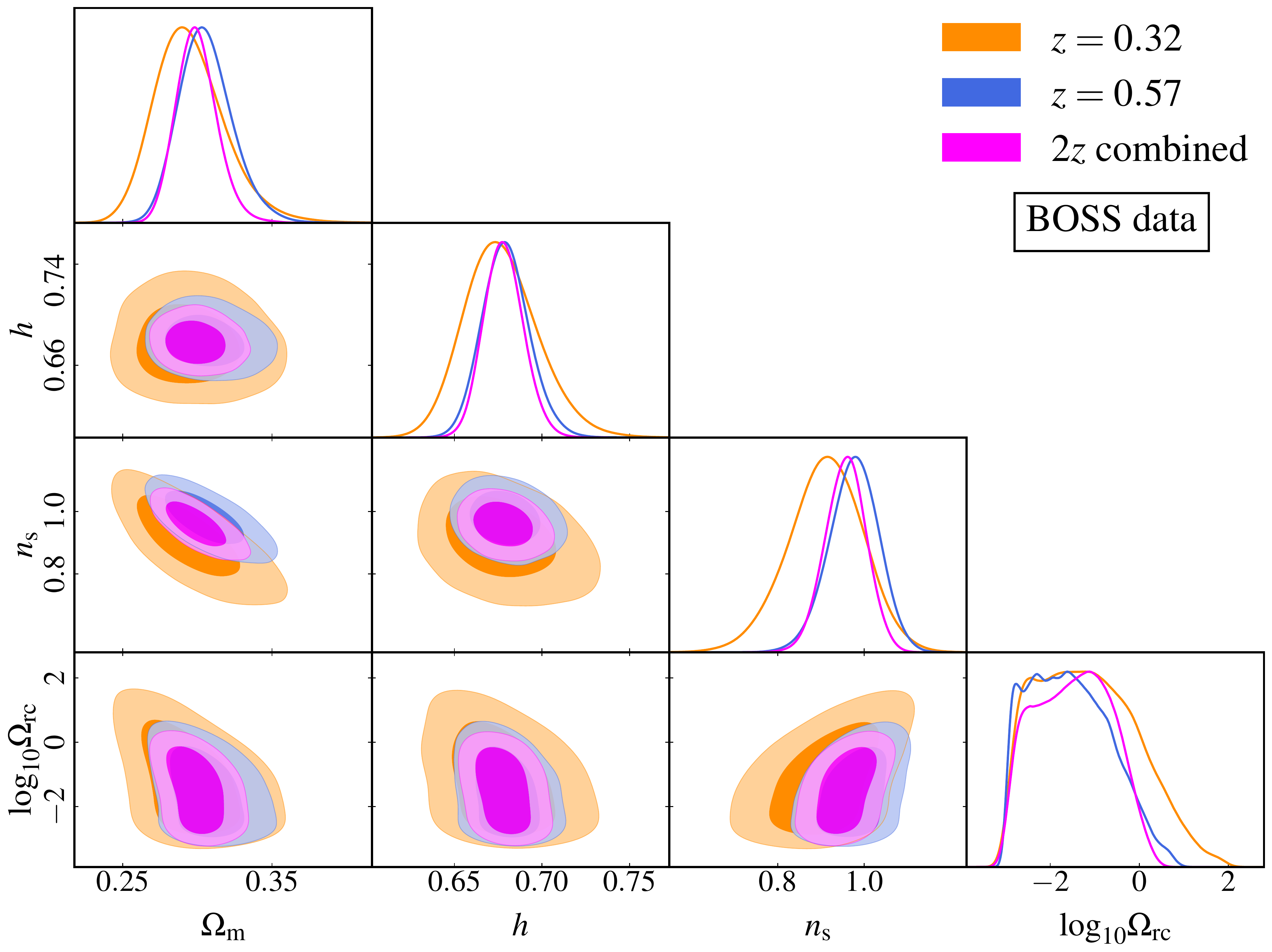}
\caption{Marginalized posteriors for the cosmological parameter from the analysis of BOSS data for all the sky-cuts with flat prior on $A_{s}$ (top panel) and $A_{s}$ fixed to  Planck central value (bottom panel), using BOSS covariances.}
\label{fig:boss}
\end{center}
\end{figure}

\begin{table}
\centering
\setlength{\tabcolsep}{0.5em}
\renewcommand{\arraystretch}{1.4}
\begin{tabular}{|l|c|c|} 
 \hline 
Param & best-fit & $\text{mean} \pm 1 \sigma$ \\ \hline 
$\omega{}_{\rm cdm }$& $0.1169$ & $0.1182_{-0.01}^{+0.0084}$   \\ 
$h$ &$0.6791$ & $0.6794_{-0.015}^{+0.013}$ \\ 
$n_{s }$ &$0.9793$ & $0.976_{-0.056}^{+0.062}$ \\
$\Omega{}_{\rm m }$ &$0.3075$ & $0.3058_{-0.02}^{+0.016}$ \\ 
\hline 
\end{tabular}
\caption{Best-fit and mean values with $1\, \sigma$ deviations  measured on BOSS data, high-$z$ sample.  
The bounds are obtained with $A_{s}$ fixed to the Planck central value and using BOSS covariances. }
\label{tab:BOSS_bounds}
\end{table}
We show the marginalized posteriors in figs.~\ref{fig:boss} and the best-fit and mean values of the cosmological parameters in tab.~\ref{tab:BOSS_bounds}. The posteriors are in qualitative agreement with those obtained from synthetic data and simulations. In particular, they show the same error size and degeneracy among parameters.

\begin{figure}[t]
    \centering
    \includegraphics[width = 0.65\textwidth]{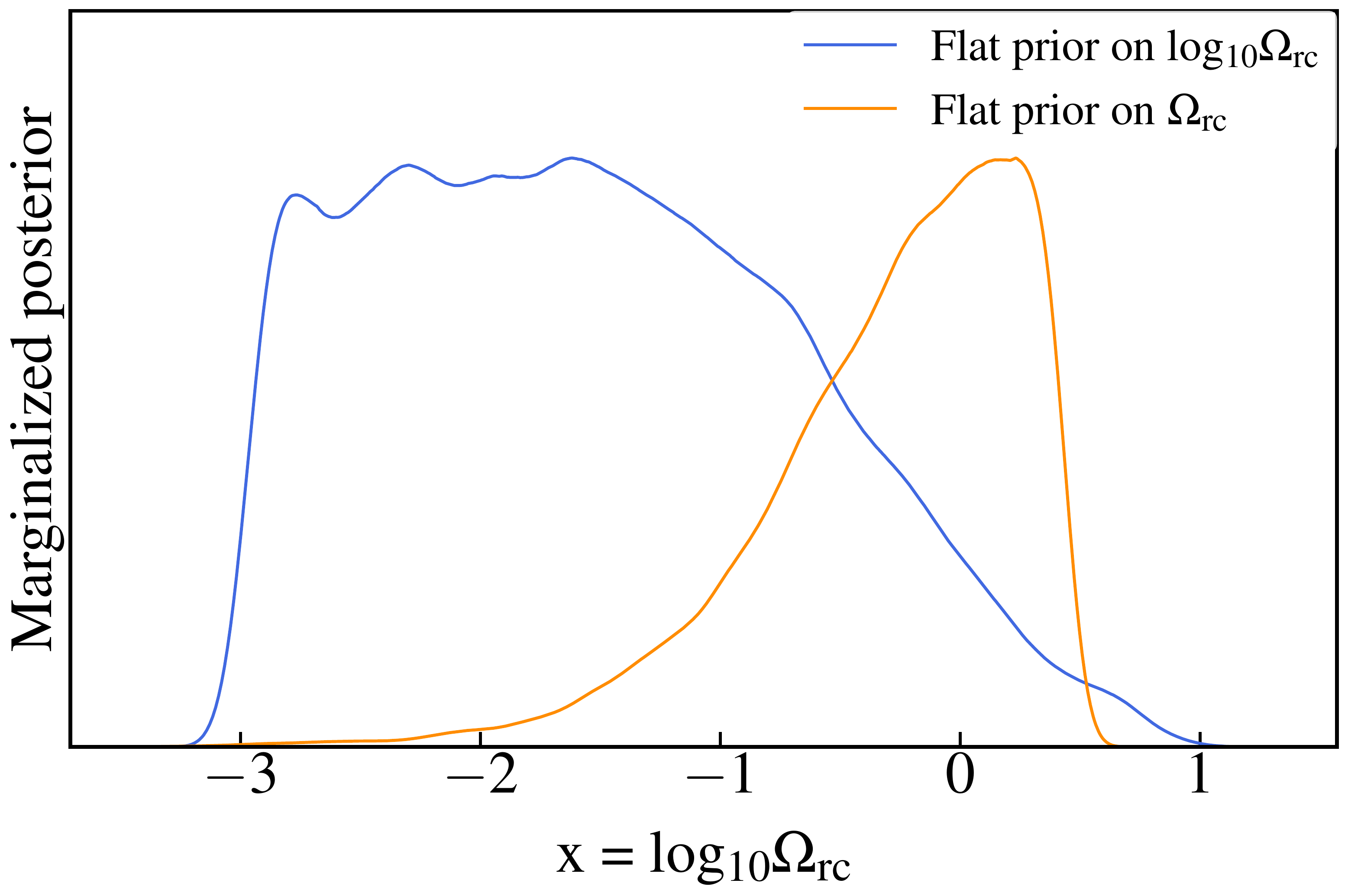}
    \caption{Marginalized posteriors of $\log_{10} \Omega_{\rm rc}$ from the BOSS analysis (high-$z$ sample only)
    with a
flat prior on $\Omega_{\rm rc}$ (yellow line) and with a flat prior on $\log_{10} \Omega_{\rm rc}$ (blue line),   with $A_{s}$ fixed to  Planck central value.}
    \label{fig:comparison_flat_priors}
\end{figure}

\subsection{Dependence on priors}

We now discuss upper bounds on the nDGP parameter $ \Omega_{\rm rc}$ from the analyses on BOSS data 
\beq
x\equiv \log_{10} \Omega_{\rm rc}\,,
\eeq
that is, ${\cal P}(x|d,{\rm p})$, where $d$  and $\rm p$ denote  the data and the priors, respectively. 
We considered two types of priors: a flat prior on $x$ (log-flat prior) on an interval $\left[-a,\,2\right]$(\footnote{The {\em upper} bounds in both priors are values of the parameters for which the posteriors are practically zero, so the results are independent on the precise choices for them.}), that is
\beq
{\rm p}^x_a(x)= {\rm flat}_{x\in\left[-a,\,2\right]}\,,
\eeq
and a flat prior on $\Omega_{\rm  rc}=10^x$ on the interval $\left[0,\,100\right]$,
\beq
{\rm p}^\Omega(x)= {\rm flat}_{\Omega_{\rm rc}\in\left[0,\,100\right]}\,.
\eeq

Due to the Jacobian of the transformation between $ \Omega_{\rm rc}$ and $x$, when expressed in the $x$ variable the second prior  is proportional to  $10^x\, \ln 10$.
For the log-flat prior, the choice of the lower bound is important.
In this case, as the $x$ parameter goes to zero, it scans an infinite volume. But since the model has a smooth limit to GR in the $\Omega_{\rm rc} \to 0$ limit, allowing a very low lower bound introduces a large parameter space in which we are effectively sampling the $\Lambda$CDM model, thus biasing the marginalized 1-d posteriors.
To remove this effect, for the log-flat prior we must choose a lower bound $a$ such that, fixing all parameters, the differences in power spectra between the case $\Omega_{\rm rc} = 0$ and the case in which $\Omega_{\rm rc} = 10^{-a}$ is comparable to the standard deviation of the data, see fig.~\ref{fig:OmrcSens}.
In fig.~\ref{fig:comparison_flat_priors} we show the 1d marginalized posteriors obtained with the flat prior on $x$, and lower bound in $x=-3$,  ${\rm p}^x_3(x)$, as in fig.~\ref{fig:boss}, and with the flat prior on $ \Omega_{\rm rc}$, ${\rm p}^\Omega(x)$. Notice that the former tends to a plateau for large negative $x$, while the latter is exponentially suppressed in the same region, given the $10^x$ dependence on the prior. These behaviors are expected: data are insensitive to small $ \Omega_{\rm rc}$ values, therefore the posterior in those region is just proportional to the prior.

We can define upper bounds on $\Omega_{\rm rc}$ from the integral,
\beq
{\cal I}[x_{\rm lim};{\rm p}]=\int_{x^p_{\rm min}}^{x_{\rm lim}} {\cal P}(x|d,{\rm p}) dx\,, 
\label{Ical}
\eeq
where $x^p_{\rm min}$ is the lower bound of the prior ${\rm p}(x)$, that is, $-3$ for ${\rm p}^x_3(x)$ and $-\infty$ for ${\rm p}^\Omega(x)$. Setting ${\cal I}[x_{\rm lim};{\rm p}]= 0.68,\,0.95$ gives the $x_{\rm lim}$ values corresponding to the  upper limits on the parameter $x$ at the $68\,\%$ or $95\,\%$ confidence level.

Translated back to $ \Omega_{\rm rc}$, these values correspond to
\begin{align}
 \Omega_{\rm rc}&< 0.0919 \quad \;{\rm at}\; 68\%\;{\rm C.L.}\;\quad(< 0.646 \;{\rm at}\; 95\%\;{\rm C.L.})\qquad\qquad {\rm for} \;{\rm p}^x_3(x)\,,\\
 \Omega_{\rm rc}&< 0.9967 \quad \;{\rm at}\; 68\%\;{\rm C.L.}\;\quad(< 2.185 \;{\rm at}\; 95\%\;{\rm C.L.})\qquad\qquad {\rm for} \;{\rm p}^\Omega(x)\,.
 \label{bound1}
 \end{align}
 These upper bounds are clearly prior-dependent. In particular, considering the flat prior on $x$, ${\rm p}^x_a$, they depend on the value of the lower extreme $x=-a$, as shifting it to $x=-b$ changes the value of the integral \re{Ical} by a quantity proportional to $b-a$, since the posterior is constant in that region. For instance, the lower extreme $x=-6$ gives 
 \begin{align}
 \Omega_{\rm rc}&< 0.0187 \quad \;{\rm at}\; 68\%\;{\rm C.L.}\;\quad(< 0.4087 \;{\rm at}\; 95\%\;{\rm C.L.})\qquad\qquad {\rm for} \;{\rm p}^x_6(x)\,.
 \label{bound_minus_6}
 \end{align}
 We stress that such a low lower bound on $x$ is not a very sensible choice.
 When calculating the bound as an integral over $x$, we are taking into account a large volume of parameter space that is simply $\Lambda$CDM.
 In the next subsection, we discuss a less prior-dependent method to extract a bound on how much we are allowed to deviate from the $\Lambda$CDM cosmology.
 
 \begin{figure}[h]
    \centering
    \includegraphics[width = 0.65\textwidth]{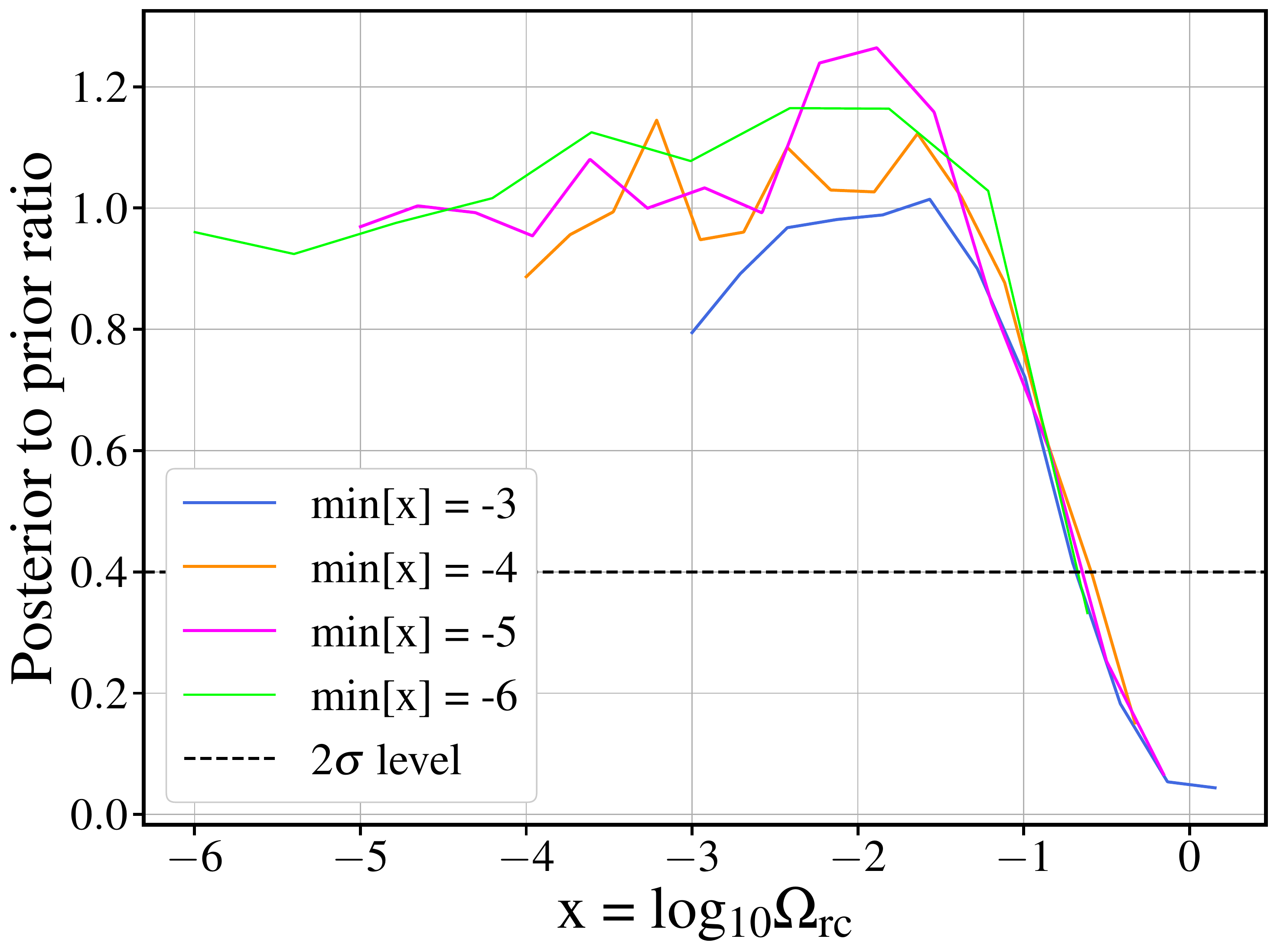}
    \caption{Normalized posterior-to-prior ratios of $\log_{10} \Omega_{\rm rc}$ from the  BOSS analysis ($2z$-combined sky-cuts), assuming a flat prior on $\log_{10} \Omega_{\rm rc}$,  with all the cosmological parameters fixed to their respective Planck central values, for different lower bounds of $\text{min} [ \log_{10} \Omega_{\rm rc}]$: $ -6, -5, -4$ and $ -3$. The dashed line denotes $p\text{-value} = 0.05$.}
    \label{fig:comparison_lower_bounds}
\end{figure}

\subsection{Bound using the Bayes factor}

 As an alternative criterion to quantify an upper bound to $ \Omega_{\rm rc}$,  consider  the ratio between the 1d posterior and its corresponding prior,
 \beq
 b(x;d,\,{\rm p})=\frac{{\cal P}(x|d,{\rm p})}{{\rm p}(x)}\,.
 \eeq
 Taking two different values of $x$ and  the ratio between the corresponding $b$ functions, gives
 \beq
 \frac{ b(x_1;d,\,{\rm p})}{ b(x_2;d,\,{\rm p})} =\frac{{\cal P}(d|x_1)}{{\cal P}(d|x_2)}\equiv B(x_1,x_2)\,,
 \eeq
 where ${\cal P}(d|x)$ is the likelihood of the data $d$ when the parameter is fixed at $x$, and the first equality is a consequence of Bayes theorem. $B(x_1,x_2)$, also known as the odds, or Bayes factor, gives the change in relative probability between the model with $x=x_1$ and $x=x_2$, supported by the data (see, for instance, \cite{Gordon:2007xm}), and is prior independent by construction. 
 
 Values of $ B(x_1,x_2)$ of order unity are  associated to `inconclusive' evidence of  $x_1$ over $x_2$. Following tab.~1 of \cite{Gordon:2007xm}, we  associate, for instance, a $p$-value of 0.05 ($2\, \sigma$) preference of $x_1$ over $x_2$ to odds in the range  $ B(x_1,x_2) > 2.5$ and a  $p$-value of 0.003 ($3\, \sigma$) to $ B(x_1,x_2) > 21$. If we fix $x_1$ to the low $x$ plateau, we can then plot $ B(x_1,x_2)$ as a function of $x_2$ and derive upper bounds on $x$. This is shown in fig.~\ref{fig:comparison_lower_bounds}, where we plot the functions $B(x_{\rm plateau},\,x)$ obtained from different MCMC chains in which we imposed flat priors on $x$, ${\rm p}^x_a(x)$, with different values for the lower bound $a$. The different curves have been normalized in order to have $b(x_{\rm plateau};d,\,{\rm p}_a^x)=1$ on the plateau, correcting by the differences between the evidences ($\int_{-a}^2 {\cal P}(d|x) p^x_a(x)$) due to the different lower limit of  integration. As we can see, while the plateau is clear for $x_{\rm min}^p= -5$ and $-6$, it is only marginally visible for higher values of $x_{\rm min}^p$. However, we also see that the different curves have a descent which is to a great extent independent on the assumed value for $x_{\rm min}^p$.
 The $1 \, \sigma$ bound extracted with the criteria described above corresponds to the tail of the distributions, which is difficult to sample numerically. Therefore, we only quote the $2\,\sigma$ bound corresponding to $B(x_{\rm plateau},x) = 2.5$, obtained from the smallest available value for $x_{\rm plateau}$, that is $-6$, and from the largest value, $-3$.
Apart from that, we use the same data and priors on the other parameters as those used to obtain fig.~\ref{fig:boss} and the bounds in \re{bound1}. 

We obtain, for $x_{\rm plateau} = -6$,
 \beq
  \Omega_{\rm rc} < 0.2113\, \quad \;{\rm at}\; 95\%\;{\rm C.L.}
 \eeq
 and, for $x_{\rm plateau} = -3$,
  \beq
  \Omega_{\rm rc} < 0.2047\, \quad \;{\rm at}\; 95\%\;{\rm C.L.}\;.
 \eeq
These two bounds are very close, showing the robustness of our approach.

\section{Conclusion}
\label{sec:concl}
We have constrained cosmological modifications of gravity with the full-shape power spectrum of BOSS data. As shown in sec.~\ref{sec:PTMG}, we have modeled the galaxy distribution in redshift space by extending perturbation theory and the bias expansion to general scale-independent modified gravity models and using the EFTofLSS  to capture the effect of the short-scale physics.

To be specific, we have then restricted our analysis  to the normal branch of the DGP scenario, assuming the background expansion of a flat $\Lambda$CDM  model. The relation between the gravitational potential and the density contrast, which in $\Lambda$CDM is given by the Poisson equation, is now modified by an enhancement of the effective Newton constant and by the presence of new non-linear terms. These effects are parameterized by a single quantity, $\Omega_{\rm rc}$. We perform the analysis with the code \code{PyBird}, that we have adapted to include the new effects (see sec.~\ref{sec:PB_nDGP}).

As discussed in sec.~\ref{sec:pipeline}, in linear theory $\Omega_{\rm rc}$ is degenerate with $A_s$ and, since we assume a flat $\Lambda$CDM background expansion, it is also degenerate with $\Omega_{\rm m}$. Furthermore, the new non-linear kernels affect the power spectrum on short scales and can be degenerate with the primordial tilt of fluctuations  $n_s$.
These strong degeneracies, in particular the one with $A_s$, shift the 1-d marginalized posteriors of the cosmological parameters from their true values.
We have quantified this shift by analyzing synthetic data generated with the EFTofLSS, which has allowed us to estimate how  degeneracy affects the marginalized posteriors through projection effects on the parameter volume.

The effect of parameter volume projection effects on the power spectrum amplitude $A_s$ has also been investigated in \cite{Simon:2022lde,Carrilho:2022mon}, showing that different choices of the priors for the EFT parameters can induce a shift on the measured $A_s$.
We have reduced this effect by fixing  $A_{s}$ to the Planck central value, combined with a BBN prior on $\omega_b$. Even when $A_s$ is fixed, there are still residual degeneracies, such as the one between $\Omega_{\rm m}$ and $n_s$.

The results of the analysis on BOSS data are discussed in sec.~\ref{sec:BOSS}. For fixed $A_s$, the marginalized posteriors are shown in  fig.~\ref{fig:boss}. The contours of the nDGP parameter $\Omega_{\rm rc}$ are prior-dependent. In order to obtain prior-independent constraints, we have considered the ratio between the 1d posterior and its corresponding prior, the so-called Bayes factor. The resulting upper bound on the nDGP parameter is $\Omega_{\rm rc} \lesssim 0.2$ at $95\%$ C.L.. This is the first measurement of the nDGP parameter performed using the full shape galaxy power spectrum from the BOSS data.
        
Our analysis shows that competitive constraints using galaxy clustering data can be  obtained only with volumes higher than BOSS  and/or more information about the parameters, either from other datasets, such as the CMB, or by narrowing the priors on the  EFT  or bias parameters~\cite{Simon:2022lde}. This could be done, for instance, by modelling the bias expansion using analytical constructs, such as the peak-background split \cite{Kaiser:1984sw,Bardeen:1985tr}, the excursion-set approach \cite{Desjacques:2010gz,Musso:2012qk} or the consistency relations of LSS~\cite{Marinucci:2019wdb, Marinucci:2020weg}, or by using more phenomenological models informed by simulation measurements.

In this respect, the bispectrum is receiving growing interest \cite{Gil-Marin:2014sta,Gil-Marin:2014baa,Gil-Marin:2016wya}. A joint analysis with the bispectrum could, in principle, help break some degeneracies, for example the one between  the growth rate $f$ (and thus $\Omega_{\rm m}$), the linear bias, $\bone$, and the primordial amplitude, $A_s$. Reference~\cite{Agarwal:2020lov} showed that by performing a joint analysis of the power spectrum and the bispectrum one could reach a $10\%$ accuracy on $f$, while \cite{Yankelevich:2018uaz} has shown that using only the bispectrum monopole significantly reduces the information content of the bispectrum, allowing only for a better estimation of the bias parameters. 
The analysis using the EFT model for the tree-level bispectrum has been recently performed on numerical simulations~\cite{Ivanov:2021kcd}, which showed an improvement of $5$-$15\%$ on the constraints on cosmological parameters. The same analysis was performed using the 1-loop bispectrum on simulations~\cite{Philcox:2022frc} and using the BOSS data~\cite{DAmico:2022osl}, showing a $\sim 10$-$30 \%$ improvement with respect to the power spectrum-only analysis.

The procedure introduced in this work can be straightforwardly extended to other models with a scale-independent growth of perturbations, such as those described by the EFT of dark  energy. Modifications of gravity can also be scale dependent if the range of the scalar force becomes of the order of the sample size, such as for instance in the Hu-Sawicki $f(R)$ model \cite{Hu:2007nk}.  Other scale-dependent effects on the growth are expected from massive neutrinos  or baryonic feedback  (see e.g.~\cite{Aviles:2021que,Parimbelli:2018yzv}). The implementation of these effects in a fast code requires more work and is left for the future.

\section*{Acknowledgments}

We are especially grateful to Tessa Baker, Ben Bose, Pedro Carrilho, Takahiro Nishimichi, Chiara Moretti and Pierre Zhang for many useful discussions related to this work.
MM acknowledges support by the Israel Science Foundation (ISF) grant No. 2562/20. FV acknowledges partial support by the ANR Project COLSS (ANR-21-CE31-0029). BSW is supported by a Royal Society Enhancement Award (grant no.~RGF$\backslash$EA$\backslash$181023). The analysis was performed on the HPC (High Performance Computing) facility of the University of Parma, whose support team we thank.

\appendix

\section{Kernels and time-dependent functions}

\label{appendixTD}

The six kernels introduced in sec.~\ref{sub2.2} in eq.~\eqref{kernels3} are defined as
			\begin{align}\label{alphakernels}
			&\alpha^1(\vq_1,\vq_2,\vq_3)=\alpha(\vq_3,\vq_1+\vq_2)\alpha_s(\vq_1,\vq_2)=O_{\alpha_s \alpha_s}(\vq_1,\vq_2,\vq_3)-\frac{1}{2}O_{\alpha_s \alpha_a}(\vq_1,\vq_2,\vq_3) \,, \\ &\alpha^2(\vq_1,\vq_2,\vq_3)=\alpha(\vq_3,\vq_1+\vq_2)\beta(\vq_1,\vq_2)=O_{\beta \alpha_s}(\vq_1,\vq_2,\vq_3)-\frac{1}{2}O_{\beta \alpha_a}(\vq_1,\vq_2,\vq_3)\,, \\
			&\beta^1(\vq_1,\vq_2,\vq_3)=2\beta(\vq_3,\vq_1+\vq_2)\alpha_s(\vq_1,\vq_2)= 2\, O_{\alpha_s \beta}(\vq_1,\vq_2,\vq_3)\,, \\ &\beta^2(\vq_1,\vq_2,\vq_3)=2\beta(\vq_3,\vq_1+\vq_2)\beta(\vq_1,\vq_2)=2\, O_{\beta \beta}(\vq_1,\vq_2,\vq_3)\,, \\
			&\gamma^1(\vq_1,\vq_2,\vq_3)=\alpha(\vq_1+\vq_2,\vq_3)\alpha_s(\vq_1,\vq_2)=O_{\alpha_s \alpha_s}(\vq_1,\vq_2,\vq_3)+\frac{1}{2}O_{\alpha_s \alpha_a}(\vq_1,\vq_2,\vq_3)\,, \\ &\gamma^2(\vq_1,\vq_2,\vq_3)=\alpha(\vq_1+\vq_2,\vq_3)\beta(\vq_1,\vq_2)= O_{\beta \alpha_s}(\vq_1,\vq_2,\vq_3)+\frac{1}{2}O_{\beta \alpha_a}(\vq_1,\vq_2,\vq_3)\,,
			\end{align}
			where $O_{\alpha_s \alpha_s}$, $O_{\alpha_s \alpha_a}$, $O_{\beta \alpha_s}$, and  $O_{\alpha_s \beta}$, are defined analogously to eq.~\eqref{OO}.
			
To shorten the following expressions, let us introduce  the following notation,
\be
M_1(a) \equiv \frac{1}{f_+( a)} \left( \frac{3 \Omega_{{\rm m},a} ( a)}{2}  \right)^2 \;, \qquad M_2(a) \equiv \frac{\nu_{22}( a)}{2}  \frac{3 \Omega_{{\rm m},a} (  a)}{2} \;.
\ee
The time dependent functions that appear in the kernels up to third order are
\begin{align}
{\cal G}_1^\lambda (a) & = \int_0^1  \left[ G_1^\lambda(a,\tilde a) f_+(\tilde a) + G_2^\lambda(a,\tilde a)  \nu_{2}(\tilde a)M_1(\tilde a)  \right] \frac{D_+^2(\tilde a)}{D_+^2( a)}  d\tilde a\,, \\
{\cal G}_2^\lambda (a) & = \int_0^1  G_2^\lambda(a,\tilde a)  \left[ f_+(\tilde a) -    \nu_{2}(\tilde a) M_1(\tilde a) \right] \frac{D_+^2(\tilde a)}{D_+^2( a)}  d\tilde a\,,
\label{Gcal}
\end{align}
and
\begin{align}
{\cal U}_1^\lambda(a) &= \int_0^1 \left\{ G_1^\lambda(a,\tilde a) f_+(\tilde a) {\cal G}_1^\delta(\tilde a) +  G_2^\lambda(a,\tilde a) M_1(\tilde a) \left[  {\nu_{2}(\tilde a)}     {\cal G}_1^\delta(\tilde a) +   M_2(\tilde a)   \right]   \right\}  \frac{D_+^3(\tilde a)}{D_+^3(a)} d \tilde a \,,\\
{\cal U}_2^\lambda(a) &= \int_0^1 \left\{ G_1^\lambda(a,\tilde a) f_+(\tilde a) {\cal G}_2^\delta(\tilde a) +  G_2^\lambda(a,\tilde a) M_1(\tilde a)   \left[  {\nu_{2}(\tilde a)}     {\cal G}_2^\delta(\tilde a)     -   M_2(\tilde a)     \right]     \right\}  \frac{D_+^3(\tilde a)}{D_+^3(a)} d \tilde a \,,\\
{\cal V}^\lambda_{11} (a)& = \int_0^1 \left\{ G_1^\lambda(a,\tilde a) f_+(\tilde a) {\cal G}_1^\theta(\tilde a) +  G_2^\lambda(a,\tilde a) M_1(\tilde a)     \left[  {\nu_{2}(\tilde a)}     {\cal G}_1^\delta(\tilde a) +   M_2(\tilde a)     \right]   \right\}  \frac{D_+^3(\tilde a)}{D_+^3(a)} d \tilde a \,,\\
{\cal V}^\lambda_{21} (a)& = \int_0^1 \left\{ G_1^\lambda(a,\tilde a) f_+(\tilde a) {\cal G}_2^\theta(\tilde a) +  G_2^\lambda(a,\tilde a)M_1(\tilde a)  \left[  {\nu_{2}(\tilde a)}     {\cal G}_2^\delta(\tilde a)      -   M_2(\tilde a)     \right]   \right\}  \frac{D_+^3(\tilde a)}{D_+^3(a)} d \tilde a \,,\\
{\cal V}^\lambda_{12} (a)& =  \int_0^1 G_2^\lambda(a ,\tilde a) \left\{    f_+(\tilde a) {\cal G}^\theta_1 (\tilde a) - M_1(\tilde a)  \left[ {\nu_{2}(\tilde a) }   {\cal G}^\delta_1 (\tilde a) +  M_2(\tilde a )  \right] \right\} \frac{D_+^3(\tilde a)}{D_+^3(a)}  d \tilde a\,,\\
{\cal V}^\lambda_{22} (a)& = \int_0^1 G_2^\lambda(a ,\tilde a) \left\{    f_+(\tilde a) {\cal G}^\theta_2 (\tilde a) -M_1(\tilde a) \left[ {\nu_{2}(\tilde a) }   {\cal G}^\delta_2 (\tilde a) -  M_2(\tilde a )  \right] \right\} \frac{D_+^3(\tilde a)}{D_+^3(a)}  d \tilde a \,.
 \end{align}
 In the previous expressions, $G_i^\lambda$, with $i = 1,2$ and $\lambda = \delta,\theta$, are the Green's functions defined by eqs.~\eqref{gdelta}--\eqref{gtheta} while $\nu_{2}$ and $\nu_{22}$ are the nDGP functions that account for the non-linear corrections of the generalized Poisson equation \eqref{sol_NL1}. Their explicit expression for the nDGP case is given by eqs.~\eqref{mu2} and \eqref{mu22}.

\section{Comparison with LSS bootstrap}
\label{app:map}

In this appendix we discuss the relation between the perturbative expansions of \cite{Donath:2020abv} used in the main text and in \code{PyBird} with the one that can be derived in the LSS bootstrap approach \cite{DAmico:2021rdb}. This approach allows one to derive the    analytic structure of the perturbative kernels of dark matter and biased tracers starting from symmetries. In this way, one can show that the expansion used in $\Lambda$CDM also applies to more general models, even in modified gravity, with the same symmetries (translational and rotational invariance, the equivalence principle, etc.) as in $\Lambda$CDM.  Indeed, here we show that the two expansions are equivalent.

\subsection{Dark matter kernels}
\label{subapp1}

Let us start by discussing the dark matter kernels.
For the matter density contrast in the bootstrap basis we have, after imposing all the symmetries \cite{DAmico:2021rdb}\footnote{In \cite{DAmico:2021rdb} we define the $n$th-order perturbation theory kernel with a $1/n!$ with respect to the standard definition. Hence, the factor $1/2$ and $1/6$ in the second- and third-order kernels in these equations. Notice also that we have changed the name of the time-dependent coefficients.}
\begin{align}
\label{F1} K_1^{(1)}(\bq_1) =& \,\,1\,,\\
\label{F2} K_1^{(2)}(\bq_1,\bq_2)  =& \,\, \frac12 \big[ 2\beta(\bq_1,\bq_2) + a_\gamma  \gamma(\bq_1,\bq_2) \big] \,,\\
 K_1^{(3)} (\bq_1,\bq_2,\bq_3)  =& \,\, \frac{1}{6}\big[ 2\beta(\bq_1,\bq_2)\beta(\bq_{12},\bq_3) + a_{\gamma\gamma}\gamma(\bq_1,\bq_2)\gamma(\bq_{12},\bq_3) \nonumber \\
&- 2 \left(a_{\gamma\alpha} - \hh\right)\gamma(\bq_1,\bq_2)\beta(\bq_{12},\bq_3) + 2 \left(a_\gamma +2a_{\gamma\alpha} - \hh \right)\beta(\bq_1,\bq_2)\gamma(\bq_{12},\bq_3)\nonumber\\
&+a_{\gamma\alpha}\gamma(\bq_1,\bq_2)\alpha_a(\bq_{12},\bq_3)  + {\rm \, cyclic} \big] \label{F3}\,,
\end{align}
where $\alpha_a(\bq_1,\bq_2) \equiv \frac{\bq_1\cdot \bq_2}{q_1^2} - \frac{\bq_1\cdot\bq_2}{q_2^2}$ and $a_\gamma$, $a_{\gamma \gamma}$, etc.. are time-dependent coefficients that depend on the  cosmological model. Analogous expressions can be given for the velocity divergence kernels. We denote the corresponding time-dependent coefficients by $d_\gamma$, $d_{\gamma \gamma}$, etc., i.e.,
\begin{align}
\label{G1} K_2^{(1)}(\bq_1) =& \,\,1\,,\\
\label{G2} K_2^{(2)}(\bq_1,\bq_2)  =& \,\, \frac12 \big[ 2\beta(\bq_1,\bq_2) + d_\gamma  \gamma(\bq_1,\bq_2) \big] \,,\\
 K_2^{(3)} (\bq_1,\bq_2,\bq_3)  =& \,\, \frac{1}{6}\big[ 2\beta(\bq_1,\bq_2)\beta(\bq_{12},\bq_3) + d_{\gamma\gamma}\gamma(\bq_1,\bq_2)\gamma(\bq_{12},\bq_3) \nonumber \\
&- 2 \left(d_{\gamma\alpha} - \hh\right)\gamma(\bq_1,\bq_2)\beta(\bq_{12},\bq_3) + 2 \left(d_\gamma +2d_{\gamma\alpha} - \hh \right)\beta(\bq_1,\bq_2)\gamma(\bq_{12},\bq_3)\nonumber\\
&+d_{\gamma\alpha}\gamma(\bq_1,\bq_2)\alpha_a(\bq_{12},\bq_3)  + {\rm \, cyclic} \big] \label{G3}\,,
\end{align}
With these definitions we find that the function $h$ in the above expressions is given by \cite{DAmico:2021rdb}
\be
\hh(a) \equiv \int^a_0 d \ln \tilde a \; f_+(\tilde a) \left[\frac{D_+(\tilde a )}{D_+(a)}\right]^2\,d_\gamma(\tilde a )\,.
\label{hfunc}
\ee

We can now compare this expansion with the one of  eqs.~\eqref{kernels1}--\eqref{kernels3} \cite{Donath:2020abv}. For instance, by comparing eq.~\eqref{F2} with eq.~\eqref{kernels2} we obtain 
\be
a_\gamma = 2\mathcal{G}_1^{\delta}\,,\quad \quad d_\gamma = 2\mathcal{G}_1^{\theta}\,.
\label{mapp2}
\ee
 Using the definitions \re{Gcal} and the equations \re{Greenf} we can verify the following relation, which enforces  the continuity equation for the second order matter kernel,
\beq
\frac{1}{f_+(a)} \frac{d}{d \ln a} {\cal G}_1^\delta(a)= 1 - 2 {\cal G}_1^\delta(a)+{\cal G}_1^\theta(a)\,,
\eeq
and can be integrated to give 
\beq
{\cal G}_1^\delta(a)=\frac{\hh(a)+1}{2}\,,
\eeq
where we have used \re{hfunc} and \re{mapp2}.

Finally, comparing the third-order kernels we have
\begin{align}
a_{\gamma\gamma} &= 2 \mathcal{U}_1^\delta +\hh - 2\mathcal{V}_{12}^{\delta}\,,\qquad
a_{\gamma\alpha} = \frac{\hh}{2} - \mathcal{V}_{12}^\delta - \mathcal{U}_1^\delta\,, \\
d_{\gamma\gamma} &= 2 \mathcal{U}_1^\theta +\hh - 2\mathcal{V}_{12}^{\theta}\,,\qquad
d_{\gamma\alpha} = \frac{\hh}{2} - \mathcal{V}_{12}^\theta - \mathcal{U}_1^\theta\,.
\end{align}
In \cite{Donath:2020abv} a tracer-independent function $Y(a)$ was introduced, defined in terms of the functions ${\cal V} ^{\theta}_{11}$ and ${\cal V} ^{\theta}_{12}$, see eq.~\eqref{kernels3}, as
\beq
\label{Ydef}
Y(a)\equiv {\cal V} ^{\theta}_{11}(a)+{\cal V} ^{\theta}_{12}(a)-\frac{3}{14}\,.
\eeq
It is related to the  function $g(a)$ by
\beq
\label{hheq}
Y(a)=\frac{g(a)}{2}-\frac{3}{14}\,.
\eeq

\subsection{Biased tracers kernels}
\label{subapp2}

For the tracer's kernels in the bootstrap approach we have \cite{DAmico:2021rdb}
\begin{align}
\Kone (\bq_1) & = \CC_{0}\,,\label{Kone}  \\
\Ktwo(\bq_1,\bq_2) & = \, \frac12 \big[ \CC_{1}+2\, \CC_{0} \,\beta(\bq_1,\bq_2)+  \CC_{\gamma_2} \, \gamma(\bq_1,\bq_2) \big] \,, \label{Ktwo} \\
\Kthree(\bq_1,\bq_2, \bq_3) &= \frac16 \big[  \CC_{2}/3 + \CC_{\gamma_3} \gamma(\bq_1,\bq_2) + 2  \CC_{1}  \beta(\bq_1,\bq_2) \nonumber \\
& +\CC_{\gamma \gamma}   \gamma(\bq_1,\bq_2)  \gamma(\bq_{12},\bq_3)   + 2\, \CC_{0} \beta(\bq_1,\bq_2)\beta(\bq_{12},\bq_3)  \nonumber \\
&+ 2 ( \hh \, \CC_{0}  -\,\CC_{\gamma\alpha} ) \gamma(\bq_1,\bq_2)\beta(\bq_{12},\bq_3)  + 2 (\CC_{\gamma_2}  + 2\,\CC_{\gamma\alpha} -  \hh \,\CC_{0})\beta(\bq_1,\bq_2)\gamma(\bq_{12},\bq_3)\nonumber\\
& + \CC_{\gamma\alpha} \gamma(\bq_1,\bq_2)\alpha_a(\bq_{12},\bq_3) + {\rm cyclic} \big] \,.
\label{generalK}
\end{align}

We wish to verify that these expressions reproduce eqs.~\eqref{Kdg1}, \eqref{Kdg2} and \eqref{Kdg3} \cite{Donath:2020abv}. This is obvious for $\Kone$ and for $\Ktwo$ if these relations hold,
\be
\label{map2}
\CC_0= \bone   \,,\qquad\, \CC_1  = - 2 (\bone - \btwo - \bfour ) \,,\qquad\, \CC_{\gamma_2}= 2 \left(\bone - \frac27 \btwo \right) \,.
\ee
To do the same comparison on the third-order kernel, it is convenient to rewrite eq.~\eqref{generalK} in the following form, 
\begin{align}
\Kthree(\bq_1,\bq_2, \bq_3) &= \frac16 \bigg[  \frac{1}{3}\, \CC_{2} + \CC_{\gamma_3} \gamma(\bq_1,\bq_2) + 2  \CC_{1}  \beta(\bq_1,\bq_2) \nonumber \\
& + 2\, \CC_{0} \beta(\bq_1,\bq_2)\beta(\bq_{12},\bq_3)  + \left[2\CC_{\gamma_2} + \CC_{\gamma a}- 2\left(\CC_{\gamma b} + \hh \CC_0\right)\right]\beta(\bq_1,\bq_2)\gamma(\bq_{12},\bq_3) \nonumber \\
& + \left(\frac{1}{4}\CC_{\gamma a} - \frac{1}{2}\CC_{\gamma b}\right) \gamma(\bq_1,\bq_2)\alpha_a(\bq_{12},\bq_3)  +  \left( \CC_{\gamma b} - \frac{1}{2}\CC_{\gamma a} + 2 \hh \CC_0 \right) \gamma(\bq_1,\bq_2)\beta(\bq_{12},\bq_3) \nonumber\\
&  +\left(\frac{1}{2}\CC_{\gamma a} + \CC_{\gamma b}\right)   \gamma(\bq_1,\bq_2)  \gamma(\bq_{12},\bq_3)+ {\rm cyclic} \bigg] \,.
\label{generalK3new}
\end{align}
To compare this expression with~\eqref{Kdg3}, we need first to replace $\vq_1 \to \vq $, $\vq_2 \to -\vq$ and $\vq_3 \to  \vk$ and then subtract the UV part, i.e.~the finite part of the kernel in the limit $q/k\to \infty$. Subtracting the UV part removes the first line of the above equation, while the term proportional to $\beta(\bq_1,\bq_2)\gamma(\bq_{12},\bq_3)$ and its permutations cancel. Moreover, also the combination of terms multiplying $\CC_{\gamma b}$ vanishes under this replacement.

Comparing what remains with eq.~\eqref{Kdg3}, using the notation eq.~\eqref{OO}, eq.~\eqref{map2} and the relation between $Y$ and $\hh$, eq.~\eqref{hheq},
we obtain
\be
\label{map3}
\bthree = \frac{21}2 \left(  \hh \CC_0    - \frac{\CC_{\gamma a}}{2} \right)\,.
\ee

Before concluding, let us compare this bias expansion with another commonly used one \cite{Eggemeier:2018qae},
\be
\delta_g = \hat b_1 \delta+ \frac{\hat b_2}{2} \delta^2 + \hat b_{{\cal G}_2} {\cal G}_2 +  \hat b_{{\cal G}_N}  {\cal G}_N + \ldots \;,  
\ee
where 
\be
{\cal G}_2 = (\nabla_i \nabla_j \varphi_1)^2 - (\nabla^2 \varphi_1)^2 \;, \qquad {\cal G}_N = \nabla_i \nabla_j \varphi_1 \nabla_i \nabla_j \varphi_2 - \nabla^2 \varphi_1 \nabla^2 \varphi_2 \;,
\ee
with $\nabla^2 \varphi_1=\delta$ and $\nabla^2 \varphi_2 = {\cal G}_2$. The ellipses denote bias operators that do not enter in the 1-loop calculation. Using the results of \cite{DAmico:2021rdb}, we find
\be
\begin{split}
\bone &= \hat b_1 \;, \quad \btwo = 
\frac72 \left[ \left(1  - \frac{a_\gamma}{2} \right) \hat b_1 +  \hat b_{{\cal G}_2} \right]\;, \\
\bthree & = \frac{ 21}{2} \left[ \left( \hh  - \frac{a_{\gamma a}}{2} \right) \hat b_1  + a_\gamma   \hat b_{{\cal G}_2} -  \hat b_{{\cal G}_N} \right]  \;, \\ 
\bfour &= - \frac12 \left[ \left( 5 - \frac72 a_\gamma \right) \hat b_1 - \hat b_2 + 7 \hat b_{{\cal G}_2}   \right]\;.
\end{split}
\ee

\section{Initial conditions in nDGP}
\label{app:IC_nDGP}

The Green's functions, eqs.~\eqref{gdelta}--\eqref{gtheta}, are defined in terms of the two independent solutions of the growth equation, eq.~\eqref{growth}.
Here we discuss how to select these two solutions in nDGP.

In general relativity, for matter domination, $\nu=1$ and $\Omega_{{\rm m},a}=\Omega_{\rm m}$,  one has two independent solutions, a growing and a decaying one, respectively
\be
    D_+(a) \propto a \;, \qquad 
    D_-(a ) \propto a^{-3/2} \;.
\ee

In  $\Lambda$CDM   one can express the two  solutions  in terms of hypergeometric functions with the above  initial conditions, see e.g.~\cite{Donath:2020abv,Lee:2009gb}.

In nDGP, one needs to solve  the differential equation for the growth function eq.~\eqref{growth} numerically. To get an idea of its solutions, let us expand it for small  $a$.
Up to order $a^{9/2}$, one finds 
\be
\frac{d^2 D}{d\ln a^2}+ \frac{1 + 3 \zeta a^3 }{2}\frac{d D}{d\ln a} - \left[ \frac32 +  \xi   a^{3/2} - \frac{3 \zeta + 4 \xi^2}{2}  a^3 + \frac{-5 \zeta \xi + 8 \xi^3}{2}  a^{9/2}  \right] D=0 \;,
\ee
where we have defined two dimensionless parameters,
\be
\zeta \equiv \frac{1 - \Omega_{\rm m}}{ \Omega_{\rm m}} \;, \qquad \xi =  \sqrt{\frac{\Omega_{\rm rc}}{\Omega_{\rm m}}} \;,
\ee
that parametrize the effect of deviating from matter dominance and general relativity, respectively. 
We notice that terms proportional to $\xi$ in the bracket start at order $a^{3/2}$ while those proportional to $\zeta$ start at $a^3$. Thus, for comparable values of $\zeta$ and $\xi$, the effect of modifying  gravity dominates over the effect of not being in matter domination.  Thus, it is crucial to include the nDGP corrections proportional to $\xi$.

We can solve the above equation perturbatively and we obtain two modes,
\begin{align}
D_+ (a) & \propto a \left( 1 + \frac{\xi}{6} a^{3/2} + \frac{-18 \zeta - 11\xi^2}{99} a^3 + \frac{-2817 \zeta \xi + 2816 \xi^2}{24948} a^{9/2} +{\cal O} (a^4)  \right) \;, \\ 
D_- (a)& \propto a^{-3/2} \left( 1 - \frac{2\xi}{3} a^{3/2} + \frac{9 \zeta - 32\xi^2}{18} a^3 + \frac{-9 \zeta \xi + 32 \xi^2}{81} a^{9/2} +{\cal O} (a^4)  \right) \;,
\end{align}
that we can call  the ``growing'' and ``decaying'', respectively, because at early times they match the usual growing and decaying solutions. However, both grow at large $a$ (\footnote{For $\xi \neq 0$, the growing of $D_- (a)$ is not an artifact of the expansion but the effect of having a time dependent Newton's constant.}) and separating the two solutions at late time requires infinite numerical precision. 

To bypass this problem in \code{PyBird}, we consider also the asymptotic behaviour of eq.~\eqref{growth} at late time, i.e.~for $a \gg a_0$, where it becomes
\be
\frac{d^2 D}{d\ln a^2}+2 \frac{d D}{d\ln a} =0\;,
\ee
with solutions
\be
    \tilde D_+(a)\propto \text{const.}\,,\qquad \tilde D_-(a)\propto \left( \frac{a_0}{a} \right)^2\,.
\ee
The second solution grows going backward in time and can be easily selected numerically, independently from $D_+$. Thus, to compute the Green's functions we use $D_+$ and $\tilde D_-$ as  two independent solutions.

\section{Additional parameter posteriors}
\label{app:APP_nDGP}

\begin{figure}
\begin{center}
\centering
\includegraphics[width=0.79\textwidth]{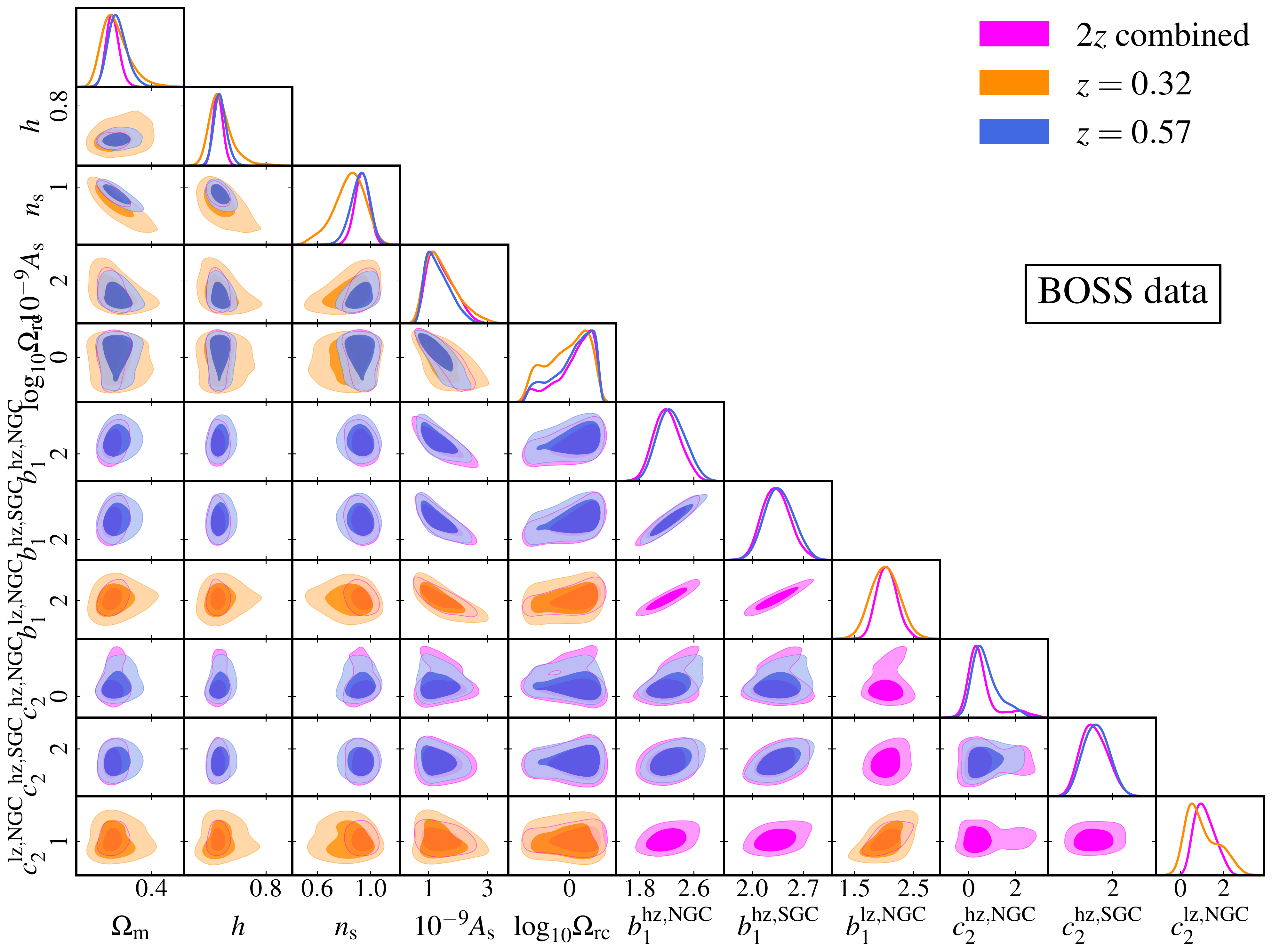}

\vspace{0.7cm}

\includegraphics[width=0.79\textwidth]{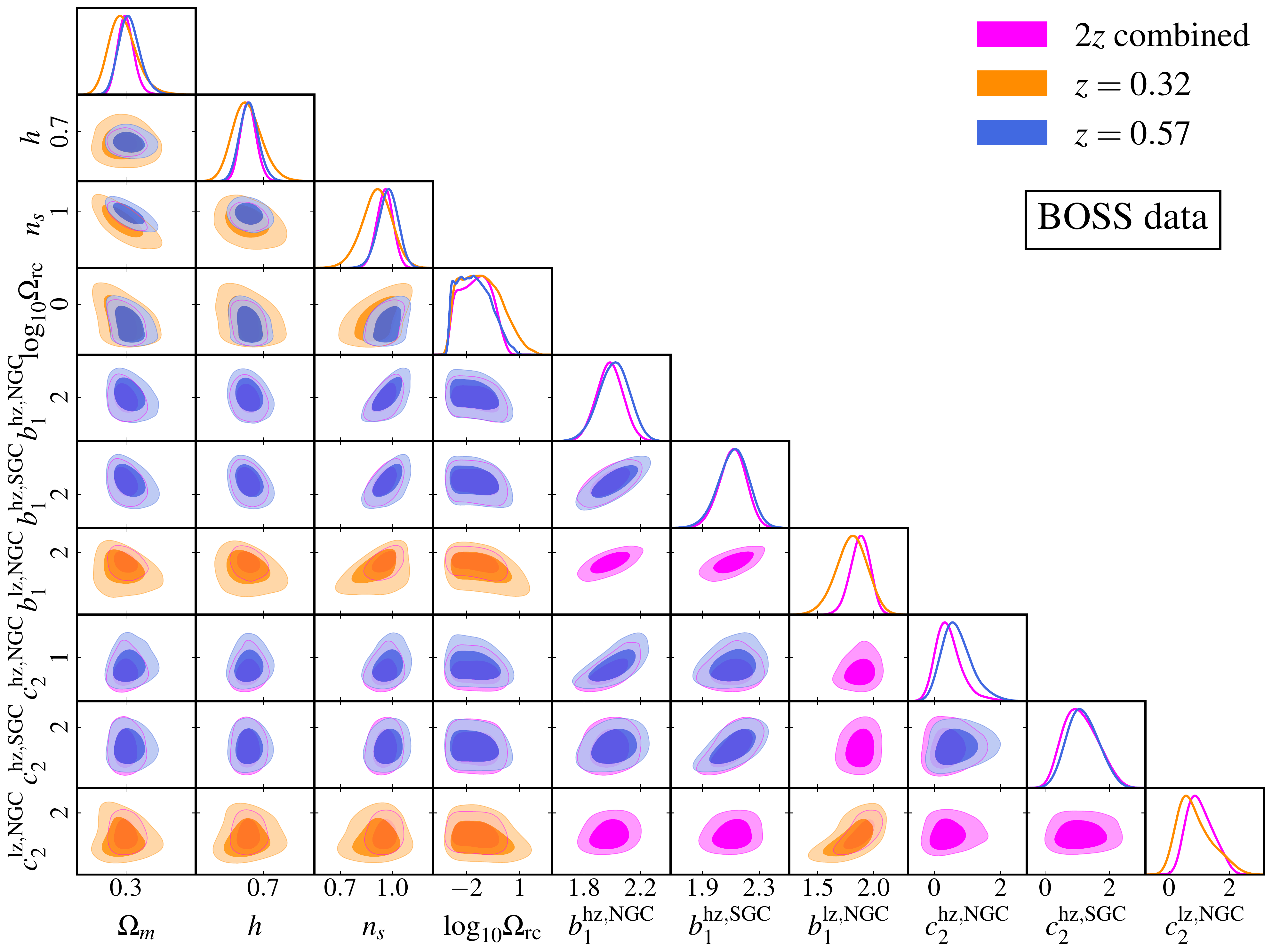}
\caption{Full triangle plots of the marginalized posteriors for the cosmological parameter from the analysis of BOSS data for all the sky-cuts with flat prior on $A_{s}$ (top panel) and $A_{s}$ fixed to  Planck central value (bottom panel), using BOSS covariances.}
\label{fig:boss_2}
\end{center}
\end{figure}

In fig.~\ref{fig:boss_2} we show the two full triangle plots obtained fitting BOSS 4 skies and including the effects of neutrinos in the linear power spectrum.

 \bibliographystyle{utphys}
\bibliography{biblio}

\end{document}